\documentclass[aps,prx,twocolumn,superscriptaddress,showpacs]{revtex4-1}
\usepackage{amsmath,amssymb,graphicx}

\usepackage[utf8]{inputenc}
\usepackage[T1]{fontenc}
\usepackage{xcolor}

\IfFileExists{newtxtext.sty}
   {\usepackage{newtxtext,newtxmath}}
   {\IfFileExists{stix.sty}
      {\usepackage{stix}}
      {\IfFileExists{mathptmx.sty}
      {\usepackage{mathptmx}}{} } }

\usepackage{textcomp}

\usepackage{bm}

\IfFileExists{siunitx.sty}{\usepackage{booktabs,siunitx}}{}

\pdfoutput=1
\usepackage{color}
\definecolor{LinkColor}{rgb}{0.256,0.439,0.588}
\usepackage{hyperref}
\hypersetup{
   pdfauthor={good guys},
   pdftitle={good title},
   colorlinks=true,
   citecolor=LinkColor,
   linkcolor=LinkColor,
   urlcolor=LinkColor
}

\renewcommand{\vec}[1]{\mathbf{#1}}
\newcommand{\bra}[1]{\langle#1\rvert}
\newcommand{\ket}[1]{\lvert#1\rangle}

\usepackage{pifont}

\usepackage{multirow}

\begin{document}

\title{Spinon Fermi surface in a cluster Mott insulator model on a triangular lattice and possible application to 1T-TaS$_2$}

\author{Wen-Yu He}
\affiliation{Department of Physics, Hong Kong University of Science and Technology, Clear Water Bay, Hong Kong, China}
\author{Xiao Yan Xu}
\email{wanderxu@gmail.com}
\affiliation{Department of Physics, Hong Kong University of Science and Technology, Clear Water Bay, Hong Kong, China}
\author{Gang Chen}
\affiliation{State Key Laboratory of Surface Physics, Department of Physics,Center for Field Theory \& Particle Physics, Fudan University, Shanghai, 200433, China}
\affiliation{Collaborative Innovation Center of Advanced Microstructures, Nanjing, 210093, China}
\author{K. T. Law}
\affiliation{Department of Physics, Hong Kong University of Science and Technology, Clear Water Bay, Hong Kong, China}
\author{Patrick A. Lee}
\email{palee@mit.edu}
\affiliation{Department of Physics, Massachusetts Institute of Technology, Cambridge MA 02139, USA}

\date{Feb 27, 2018}

\begin{abstract}
1T-TaS$_2$ is a cluster Mott insulator on the triangular lattice with 13 Ta atoms forming a star of David cluster as the unit cell. We derive a two dimensional XXZ spin-1/2 model with four-spin ring exchange term to describe the effective low energy physics of a monolayer 1T-TaS$_2$, where the effective spin-1/2 degrees of freedom arises from the Kramers degenerate spin-orbital states on each star of David. A large scale density matrix renormalization group simulation is further performed on this effective model and we find a gapless spin liquid phase with spinon Fermi surface at moderate to large strength region of four-spin ring exchange term. All peaks in the static spin structure factor are found to be located on the "$2k_F$" surface of half-filled spinon on the triangular lattice. Experiments to detect the spinon Fermi surface phase in 1T-TaS$_2$ are discussed.

\end{abstract}

\maketitle

Quantum spin liquid (QSL) was first proposed by P. W. Anderson in 1973~\cite{Anderson1973}. He argued 
that the ground state of spin-1/2 Heisenberg antiferromagnet on the triangular lattice  
is a random quantum superposition of singlets, the so called resonating valence bonds (RVB). 
Although the RVB state is not the true ground state of the triangular lattice spin-1/2 
Heisenberg model, Anderson's proposal has inspired a great deal  of study this new "quantum liquid" state in frustrated magnetic systems.
QSL is a highly entangled states and is very difficult to realize and characterize in experiments
due to the lacking of an obvious order parameter and symmetry breaking.
During the past forty years people only find a few  QSL candidates, such as organic compounds $\kappa$-(BEDT-TTF)$_2$Cu$_2$(CN)$_3$~\cite{Shimizu2003} and EtMe$_3$Sb[Pd(dmit)$_2$]$_2$~\cite{Itou2007}, herbertsmithite (ZnCu$_3$(OH)$_6$Cl$_2$)~\cite{Shores2005}, Na$_4$Ir$_3$O$_8$~\cite{Okamoto2007}, YbMgGaO$_4$~\cite{Li2015a,Li2015b,shen2016evidence,paddison2017continuous} and recently proposed 
1T-TaS$_2$~\cite{Law2017,Klanjvsek2017,Yu2017,Ribak2017}, but still with many controversies in details. 
The ongoing efforts are either in the direction to explore new QSL candidate materials~\cite{Feng2017,Ding2018}, 
or push our theoretical and numerical understanding further.

It is well-known that, the geometrically frustration on kagome, pyrochlore and triangular lattices, or
spin anisotropy such as Kitaev type interaction on a honeycomb lattice~\cite{Kitaev2006}, play an important role to stabilize a QSL phase~\cite{Balents2010}. On a kagome lattice, the isotropic nearest neighbor antiferromagnetic Heisenberg interaction  is probably enough to result in QSL phase based on density matrix renormalization group (DMRG)~\cite{Yan2011,Depenbrock2012,Jiang2012,Liao2017} or variational Monte Carlo (VMC)~\cite{Iqbal2013} calculations, while on a triangular lattice, it is not the case. The ground state of Heisenberg model on the triangular lattice is the $120^\circ$-AFM state~\cite{Satoru1992,Chubukov1994,White2007}. Thus to stabilize QSLs, more frustration, such as next neighbor frustrations~\cite{Li2015,Zhu2015,Hu2015,Ryui2014,Iqbal2016}, anisotropic~\footnote{there is spatial anisotropy in organic compounds $\kappa$-(BEDT-TTF)$_2$Cu$_2$(CN)$_3$~\cite{Shimizu2003} and EtMe$_3$Sb[Pd(dmit)$_2$]$_2$~\cite{Itou2007}} or high order exchange interactions is needed. The ring exchange terms become important for systems close to the insulating side of the Mott transition and it is suggested that the organics belongs to this case~\cite{Motrunich2005, Lee2005}. Exact diagonalization and variational study of the triangular lattice spin model with ring exchanges find 
a gapless QSL ground state with a spinon Fermi surface~\cite{Motrunich2005}.
Later DMRG simulation on two and four spin ladders and Gutzwiller variational wave functions 
calculation also find a similar QSL phase~\cite{Sheng2009,Block2011}.

1T-TaS$_2$ was proposed to be a QSL candidate by two of us~\cite{Law2017}. It 
has quasi-2D structure and each layer is made up of a triangular lattice with Ta atoms. 
It is recognized
that 13-site clusters are formed with
very narrow band near Fermi surface due to spin-orbit coupling (SOC)~\cite{Fazekas1979,Rossnagel2006}. A weak residual repulsion interaction is enough to open a Mott gap. Charge fluctuations  induce  high order exchange processes 
for the local moments if the system is close to the  Mott transition (a weak Mott insulator). There are good reasons to expect this to be the case for 1T-TaS$_2$ because it is the only insulator among all the CDW compounds and a related material 1T-TaSe$_2$ is metallic.  This motivates us to derive an effective spin model that include the effect of SOC and ring exchange. 
The geometric frustration and high order exchange interaction and spin anisotropy
are new ingredients for the possible QSL physics in 1T-TaS$_2$.
In this paper 
we first derive a microscopic effective spin model including the anisotropy modified ring exchange interaction for this kind of material, and then perform the state-of-art large-scale DMRG simulation to explore ground state over quite a large range of parameter space. Our work will not only shed a new light in the understanding of ground state of 1T-TaS$_2$, but will also push the limit of DMRG results for XXZ model with ring exchange on the triangular lattice, which is relevant to many other materials.

{\it Effective spin model of 1T-TaS$_2$}\,---\,
In 1T-TaS$_2$, the Ta atoms form a planar triangular lattice sandwiched by S atoms in an octahedral coordination. The Ta layer and S layers have the ABC type stacking, which restores the global inversion symmetry for the crystal structure. As the temperature is lowered, 1T-TaS$_2$ undergoes a series of charge-density wave phase (CDW) transition
and eventually entering the commensurate CDW phase around 180K. This  is  the Mott insulating state~\cite{sipos2008mott} where the lattice is deformed into a superlattice with the unit cell of star of David, formed by 13 Ta clustered atoms seen from Fig.~S1 (b) of supplementary materials (SM)~\cite{suppl}. In the $\sqrt{13}\times\sqrt{13}$ star of David unit cell, the outer twelve Ta atoms have displacement toward the centered Ta atom, which strengthens the interatomic bonds inside the star of David and weakens others. As the early first principles bulk band structure calculation for 1T-TaS$_2$ indicates that the Ta $5d$ orbitals are dominant in the conduction and valence bands~\cite{Mattheiss1973,Smith1985,Rossnagel2006,Darancet2014,Qiao2017}, the atomic SOC from $d_{x^2-y^2}$ and $d_{xy}$ orbitals is expected to modify the reconstructed band structure in the commensurate CDW phase. Importantly, the joint effect of lattice deformation and atomic SOC gives rise to the well isolated narrow band at the Fermi level, as is shown in Fig.~S1 (a) of SM~\cite{suppl}. 
As a result, in the presence of weak repulsive interaction, the 1T-TaS$_2$ is susceptible to the Mott-Hubbard transition and turns out to be a Mott insulator.

\begin{figure}[t!]
\includegraphics[width=0.9\columnwidth]{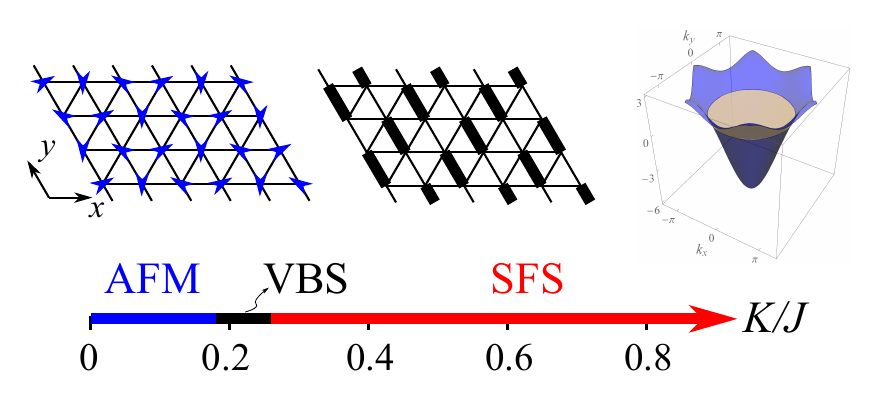}
\caption{Phase diagram  for isotropic case ($\gamma=0$), while for small anisotropy case related to real materials 1T-TaS$_2$  the phase diagram is similar. It is mainly obtained from six wide systems and confirmed in eight wide systems. Here AFM denotes $120^\circ$-spin order; VBS denotes valence bond solid state (or call dimerized phase); SFS denotes a quantum spin liquid with a spinon Fermi surface.
}
\label{phase-diagram}
\end{figure}

In order to describe the Mott state in the 1T-TaS$_2$, we consider a single star of David unit cell as a super-site, which is described by the intra-cluster tight binding Hamiltonian. Through numerical diagonalization, the Wannier orbitals localized inside the star of David with corresponding eigen-energies can be obtained in terms of the linear combination of atomic orbitals from the 13 Ta atoms. At the energy of the narrow band, it is found that the Wannier orbitals $\Psi^{\uparrow}_{\alpha}$ and $\Psi^{\downarrow}_{\beta}$ form the Kramers doublet while the Wannier orbitals $\Psi^{\downarrow}_{\alpha}$ and $\Psi^{\uparrow}_{\beta}$ are lifted in the energy due to the atomic SOC. Here the expressions for the two Wannier orbitals can be found in SM~\cite{suppl}. Taking the two Wannier orbitals as the basis, we can construct a two-orbital Hubbard model with both inter-orbital and intra-orbital interactions for 1T-TaS$_2$~\cite{suppl}. Since each star of David unit cell occupied with the single state $\Psi^{\uparrow}_{\alpha}$ or $\Psi^{\downarrow}_{\beta}$ would have the lowest energy, all other occupation states can be perturbatively dealt with through the Schrieffer-Wolff transformation. As a result, the effective XXZ spin model with the anisotropy modified ring exchange terms can be  obtained as
\begin{widetext}
\begin{align}\nonumber
\tilde{\mathcal{H}}_{\textrm{eff}}&=J\sum_{\left \langle i,j \right \rangle}\left(S_i^x S_j^x+S_i^yS_j^y+(1+\gamma) S_i^z S_j^z\right)  +K\sum_{\left \langle i, j, k, l \right \rangle}\left[{\left(S_i^x S_j^x+S_i^y S_j^y+(1+\gamma) S_i^z S_j^z\right)\left(S_k^x S_l^x+S_k^y S_l^y+(1+\gamma) S_k^z S_l^z\right)}\right.\\
&\left.{+\left(S_j^x S_k^x+S_j^y S_k^y+(1+\gamma) S_j^z S_k^z\right)\left(S_i^x S_l^x+S_i^y S_l^y+(1+\gamma) S_i^z S_l^z\right)-\left({\bm S}_i\cdot{\bm S}_k\right)\left({\bm S}_j\cdot{\bm S}_l\right)}\right].
\label{Heff}
\end{align}
\end{widetext}
In the effective spin model, the $J$-term is a XXZ type nearest neighbor interaction, where $\gamma$ denotes spin anisotropy, which arises as the ratio between the inter-orbital and intra-orbital interaction deviates from one~\cite{suppl}.   Due to the atomic SOC, the effective spin model does not have the SU(2) spin rotational symmetry but preserves the U(1) rotation around the $z$ direction. Eq.~\eqref{Heff} is of general interest as an effective spin Hamiltonian including SOC. Therefore, given the large SOC in 1T-TaS$_2$, the smallness of $\gamma$ was not obvious a priori and required a demonstration. However in practice, it turns out that for 1T-TaS$_2$ when the inter-orbital and intra-orbital interactions are in the same order, the anisotropy $\gamma$ remains smaller than 0.1~\cite{suppl}. In the large limit of atomic SOC, the anisotropy $\gamma$ will be further suppressed~\cite{suppl}.  In the rest of the paper we will mostly treat the case $\gamma$=0. The $K$-term is the four spin ring exchange term and is modified by the spin anisotropy. In general, the strength of $K/J$ depends on 
the ratio between the effective in-plane hopping and interaction. In the weak Mott insulating regime, the effective hopping and interaction are at the same scale, which is verified in several first principle calculations of 1T-TaS$_2$~\cite{Darancet2014,Qiao2017}, and then the strength of $K/J$ is of order one.
The details on the derivation of the effective spin model 
and a comparing of parameters definition with  earlier studied ring exchange model~\cite{Motrunich2005,Sheng2009,Block2011} can be found in 
SM~\cite{suppl}.

\begin{figure}[t]
\includegraphics[width=0.9\columnwidth]{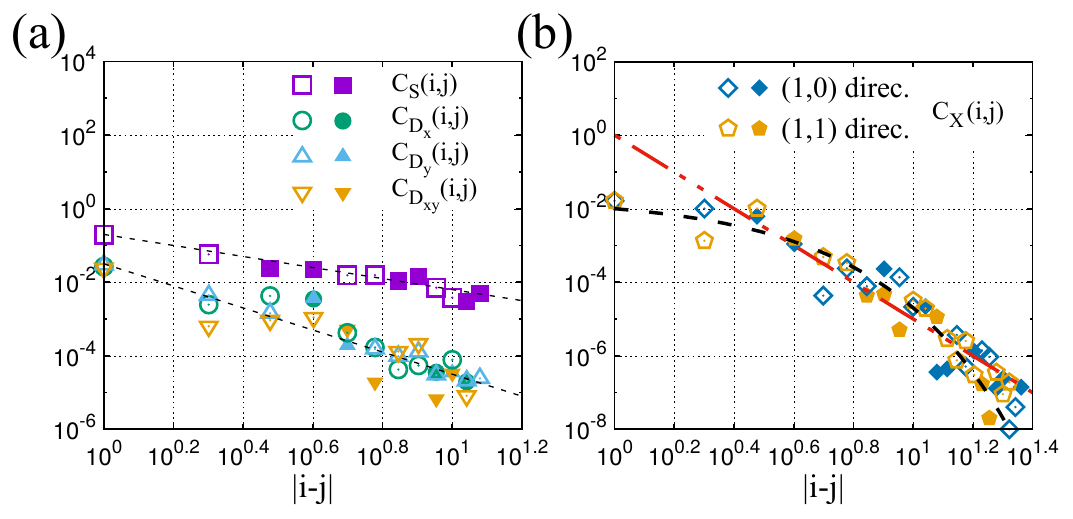}
\caption{(a) Real space decay of spin spin and dimer dimer correlation, along (1,0)-direction. The absolute values are plotted, with open symbols and filled symbols denoting that the original values are negative or positive respectively. The parameters are $L_x=24$, $L_y=6$, $K/J=0.6$, $\gamma=0$. In the log-log plot, all the correlation decays can be fitted with power law. Two power law lines are plotted to guide the eyes, top black dashed line is $|i-j|^{-1.5} $, bottom black dashed line is $|i-j|^{-3}$. (b) Real space decay of chirality chirality correlation along (1,0) and (1,1) directions with same parameters. It can be well fitted by either a high power law $|i-j|^{-5}$ (red dashed line) or an exponential decay $10^{-0.3|i-j|}$ (black dashed line).}
\label{corr_decay}
\end{figure}

For the spin model in Eq.~\eqref{Heff}, there are some well-known limits.
 (i) \textit{$K/J=0$, $\gamma=0$ case}. 
In this case, we have a pure Heisenberg model on the triangular lattice
and the ground state is the famous $120^\circ$-AFM state~\cite{Satoru1992,Chubukov1994,White2007}. 
(ii) \textit{$K/J=0$,$\gamma \rightarrow \infty$ case}.
When $\gamma=\infty$, we have a pure Ising model 
on the triangular lattice. Due to the geometry frustration, 
the Ising spin does not order at zero temperature. As this 
paramagnetic state is highly degenerate, small perturbation 
may drive it to an ordered state
via the order by disorder~\cite{villain1980order}. 
(iii) \textit{$K/J=\infty$, $\gamma=0$ case}.
In this case, we only have isotropic four-spin exchange terms. 
The ground state  in the classical limit has been discussed in Ref.~\cite{Kubo1997}. As in real materials, $K$ is usually in the same order of $J$ or smaller, this case is less relevant.

{\it Results}\,---\,
For general values of $K/J$ and $\gamma$, the ground states are not known. To identify 
all possible ground states over a wide range of parameter space, we use DMRG to solve
 the effective spin model~\eqref{Heff}. The
matrix product states (MPS) representation is used in our DMRG simulation. Due to the
$U(1)$ spin rotational symmetry, the model has a total $S^z$ conservation, and all results
 are obtained in $S_\text{tot}^z=0$ sector~\footnote{Except the calculation with magnetic fields, which is performed with general DMRG}. We use the cylindrical geometry with open boundary condition in the $x$ direction (See  Fig.~S4 of SM~\cite{suppl}). We use  a bond dimension up to  $m=5120$ and all results are obtained with a truncation error 
 in or less than the order of $10^{-5}$. 

To detect possible orders, we measure the spin-spin correlation $\langle\vec{S}_{i}\cdot\vec{S}_{i'}\rangle$,
the dimer-dimer correlation $\langle D_{b}(i)D_b(i')\rangle-\langle D_b\rangle^{2}$ 
where $D_{b}(i)=\vec{S}_{i}\cdot\vec{S}_{i+\delta}$ (where $b=x,y,xy$ denotes $\delta = \hat{x}$, $\hat{y}$, or, $\hat{x}+\hat{y}$),
the chirality-chirality correlation $\langle X_{\triangle}X_{\triangle'}\rangle$ where
$X_{\triangle}=\vec{S}_{i}\cdot\left(\vec{S}_{j}\times\vec{S}_{k}\right)$ (with $i,j,k\in\triangle$). The phase diagram 
in
Fig.~\ref{phase-diagram} shows the results that we obtain mainly from six wide systems ($L_y=6$) and confirm in eight wide systems ($L_y=8$). Our work extends previous results done on four wide systems ~\cite{Sheng2009,Block2011}. The details of finite size scaling to obtain the phase diagram can be found in SM~\cite{suppl}.  It turns out the small anisotropy ($\gamma \lesssim 0.1$) only shifts phase transition point slightly, thus we mainly focus on the isotropic case ($\gamma=0$). At small $K/J$, the ground state is the $120^\circ$-AFM state.  In the intermediate value of $K/J$, a staggered valence bond solid (VBS) phase emerges. When $K/J>0.3$, we enter a  QSL state with a spinon Fermi surface(SFS). This state was called a spin Bose metal in some earlier literature ~\cite{Sheng2009,Block2011}

\begin{figure}
\includegraphics[width=0.9\columnwidth]{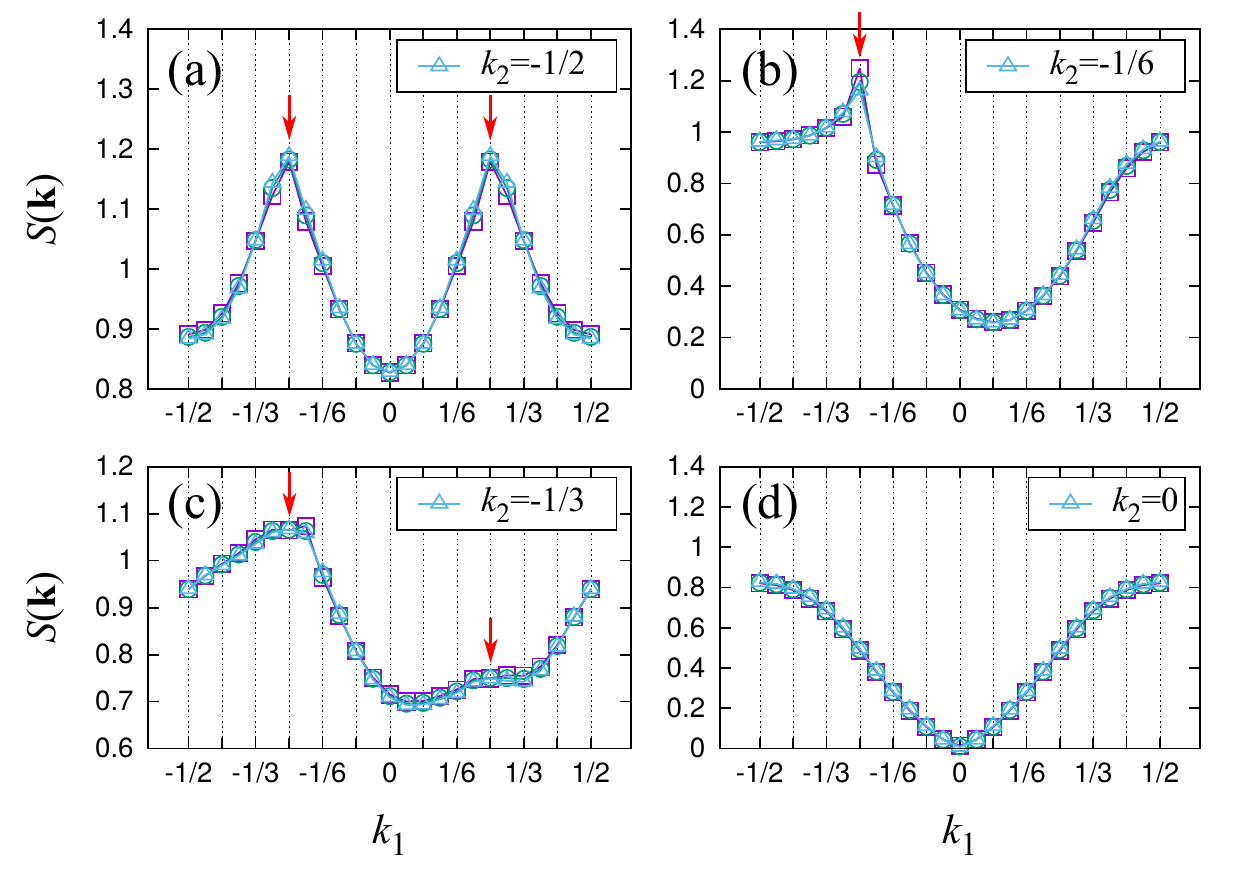}
\caption{Spin structure factor  for different $k_2$ lines. Square, circle and triangular points are results of $m=1280$, $2560$ and $5120$ respectively, where the overlap of those three points indicate the well convergence of spin structure factor. Here the parameter is $L_y=6$, $L_x=24$, $K/J=0.8$, $\gamma=0$, in SFS phase.  The red arrows denote peaks. The positions of peaks are $(\pm \frac 1 4, -\frac 1 2)$ in (a); $(-\frac 1 4,-\frac 1 6)$ in (b); $(\pm\frac 1 4, -\frac 1 3)$) in (c). For $k_2>0$, there are also peaks, the positions are inversion images of above points. }
\label{skpeaks}
\end{figure}

We here focus on the SFS phase.  In the SFS phase, all structure factors, including the spin, dimer and chirality, shows no features of ordering.  Thus we can rule out all spin, dimer and chirality orders in this phase. 
Second, the real space correlations of spin-spin and dimer-dimer  show long range correlations 
and generally can be well fitted with power laws, as showed in Fig~\ref{corr_decay}.
The long range correlation is consistent with a gapless phase rather than a gapped one. 
In addition to a global power law decay in all these correlation functions, there are modulations and sign changes superposed over them. In the following, we will see the modulations in the spin correlator are
actually  the consequence of "$2k_F$" singularity due to the existence of a spinon Fermi surface. The modulation of the dimer correlator is discussed further in the SM~\cite{suppl}. Thus the scatter of the data in Fig.~\ref{corr_decay} is due to these modulations and not due to numerical noise. As expected for a six-spin correlator, the chirality-chirality correlator shows much more rapid decay and can be fitted either with a large power law or an exponential. In addition, it exhibits sign changes. Thus we cannot interpret the chirality correlator in terms of the gauge flux correlator which is expected to have a power law decay with no sign changes in the asymptotic long distance limit. Apparently the system size is too small to reach that limit.

Another evidence for a gapless spin liquid comes from the finite static spin susceptibility.
 As shown in the SM~\cite{suppl}, we apply a small magnetic filed $B$ along $z$-direction, measure magnetic moment density $M$ along the same direction and calculate the static spin susceptibility by $\chi=\frac {\partial M_z} {\partial B_z}$. To reduce the finite size effect, several twisted boundary conditions are considered. We find that the magnetic moment density and magnetic fields obey a linear behavior and we get a finite static spin susceptibility ($\chi \approx 0.22$ for $K/J=0.8$ where $J$ and $g\mu_B$ are set to one).
Based on the value of the finite static spin susceptibility and using a half-filled free spinon band theory (static spin susceptibility is predicted to be $(g\mu_B)^2 N(0)/4$, where density of states  at Fermi surface is about $N(0)\approx\frac {1} {1.65\pi t}$ with $t$ the spinon hopping parameter), we can estimate the spinon hopping $t$ to be $\sim 0.22J$. This estimate is not precise because we expect Fermi liquid corrections to the free spinon expression for the susceptibility. The important point is that the finite spin susceptibility is consistent with a gapless spin liquid with a spinon Fermi surface. 

\begin{figure}[t]
\includegraphics[width=0.9\columnwidth]{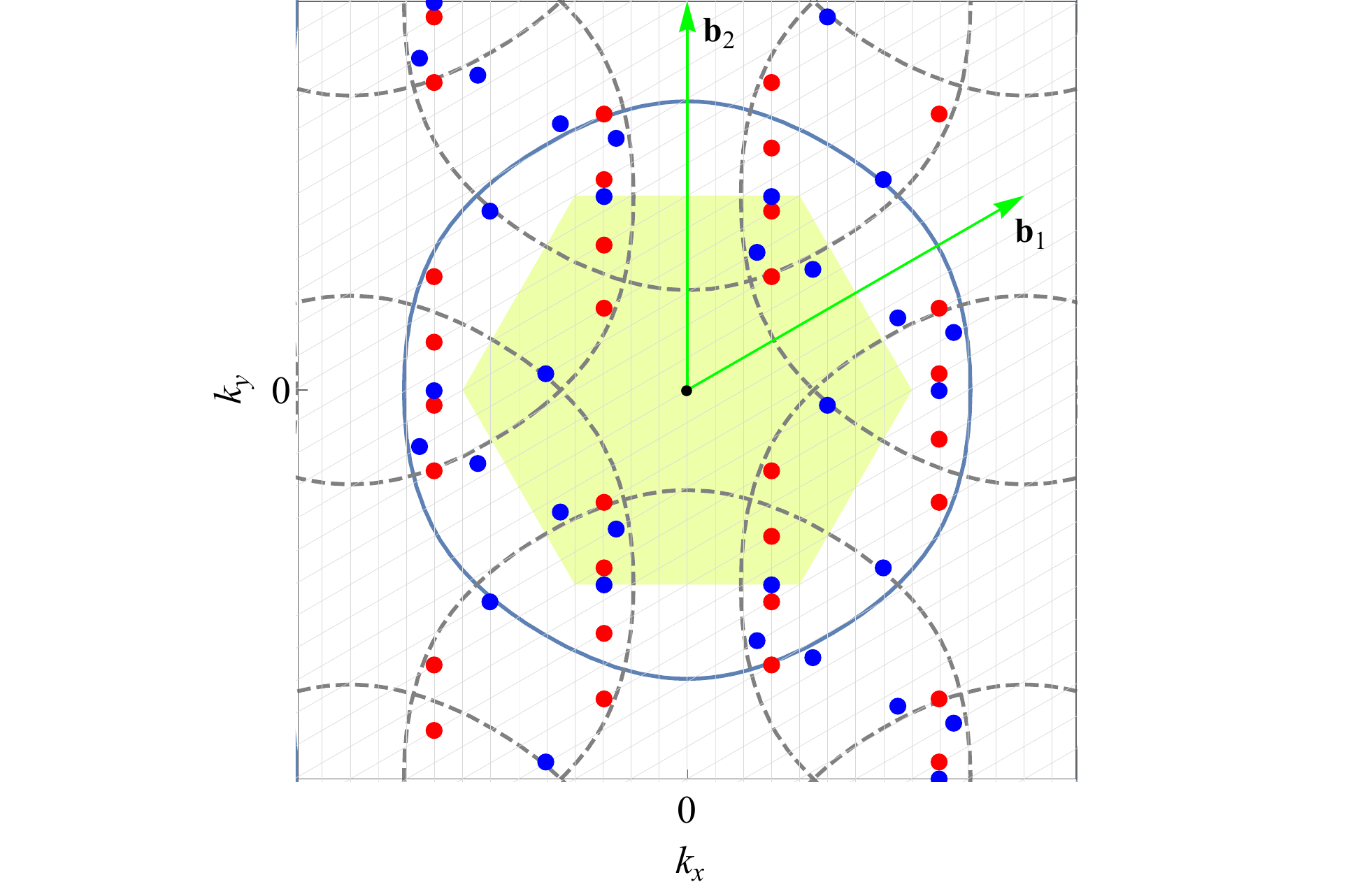}
\caption{Positions of peaks in spin structure factor with parameter $K/J=0.8$, $\gamma=0$, in SFS phase. The red points are from $L_y=6$, $L_x=24$ system, and the blue points are from $L_y=8$, $L_x=24$ system.  The solid blue line is the "$2k_F$" surface line, with $k_F$ is the Fermi vector of half-filled free spinon on the triangular lattice. Dashed gray lines are got from translation of $2k_F$ along $\vec{b}_1$, $\vec{b}_2$ and $\vec{b}_1-\vec{b}_2$. Grid size of gray mesh corresponds to $\frac 1 {12}$ of the reciprocal lattice vector. Light yellow shadowed region denotes the first Brillouin zone.}
\label{sfs}
\end{figure}

In order to confirm the SFS state, we look for $2k_F$ peaks in the spin structure factor $S(\vec{k})$. To simplify the discussion, we write $\vec{k}$ in the basis of reciprocal lattice primitive vectors $\vec{b}_1$ and $\vec{b}_2$, namely, $\vec{k}=k_1 \vec{b}_1 + k_2 \vec{b}_2$. We analyze the spin structure factor $S(\vec{k})$ for each fixed $k_2$ line, and pick out all peaks, as is shown in Fig~\ref{skpeaks}. The positions of all those peaks are plotted in Fig.~\ref{sfs}. Here results of  both $L_y=6$ and $L_y=8$ wide systems are plotted together. All the points are located on the $2k_F$ surface of a half-filled spinon on the triangular lattice, within finite size resolution. This strong signature of the existence of spinon Fermi surface is a definitive evidence for SFS phase.

We point out that there is an unexpected feature of the spin structure factor, namely there is no peak at the $2k_F$ position along the $\vec{b}_1$ line. This is seen in Fig.~\ref{skpeaks}(d) ($k_2=0$ curve) and agrees with earlier results on four wide systems~\cite{Block2011}. 
A possible explanation is that there are residual antiferromagnetic spin fluctuations associated with the $120^\circ$-spin order located at the K point $(\frac 1 3, \frac 1 3)$.  These may connect the Fermi surface crossings near $(\frac 1 3, 0)$ and $(0, -\frac 1 3$) and suppress the density of states there. 
An even more interesting possibility is that a gap is open along the $\Gamma$ (0,0) to M ($\frac 1 2,0$) directions due to spinon pairing. The Amperean scenario proposed by Lee et al~\cite{lee2007amperean} will create a set of gaps separated by spinon Fermi arcs. In this case the spin liquid will belong to the class of Z2 QSL with spinon Fermi surface. Further discussion of this possibility and its relation to the structure of the  dimer correlator can be found in the SM~\cite{suppl}.

The partially filled spinon Fermi surface have an infinite number of gapless modes, but on a finite width cylinder, there are only finite numbers of them. A fitting of central charge $c$ has been done by analyzing the entanglement entropy calculated with DMRG, which measures the number of gapless modes and is found to increase roughly linearly with width of cylinder, consistent with SFS phase. 
As discussed in the SM~\cite{suppl},the central charge can in principle distinguish between the U(1) and Z2 QSL. Unfortunately in the SFS phase the entanglement entropy converges slowly when we increase the number of states in DMRG, and it is not possible to extract the central charge unambiguously.  On the other hand the positions of peaks as well as the spin structure factor  converge faster, as showed in Fig~\ref{skpeaks}.

Another interesting finding is the VBS phase. As shown in the phase diagram (Fig.~\ref{phase-diagram}), at intermediate $K/J$, there is a VBS phase that is revealed by the dimer-dimer correlation. We clearly observe y-direction dimer structure with sharp peak at $M$ point ( $(\frac 1 2, \frac 1 2)$ in the basis of reciprocal lattice primitive vectors). Detailed results can be found in SM~\cite{suppl}. From the continuous dependence of energy on $K/J$ in our DMRG simulation on finite sizes, it is very likely that the transition from AFM to VBS or VBS to SFS is
a continuous one, where the AFM to VBS transition may generate a deconfined  quantum critical point (DQCP)~\cite{Senthil2004} and worth a further study.

As we have previously mentioned, the anisotropy in 1T-TaS$_2$ is actually small. 
We have explored the effect of this small anisotropy to the phase diagram, and found that 
the small anisotropy moderately suppresses the region of VBS phase
and thus helps stabilize the QSL phase.

{\it Discussion and Conclusions}\,---\, 
We have derived the effective spin model for 1T-TaS$_2$. The effective model is essentially an 
XXZ model with a four-spin ring exchange term on the triangular lattice. We have demonstrated that the effective model supports a SFS ground state at moderate to large $K/J$ regions. The SFS 
that we find is a gapless QSL phase with a spinon Fermi surface. We clearly observe 
the singular wavevectors in the spin structure factor from the "$2k_F$" surface that can be  described by half-filled spinon with a uniform hopping on the triangular lattice. 

Finally we remark on the applicability of these results to 1T-TaS$_2$. Our monolayer model is directly applicable to mono-layer 1T-TaS$_2$ which can be grown by MBE. For bulk samples, if the interlayer coupling is weaker than the intra-layer exchange $J$, it is possible that the ground state is made up of weakly coupled layers of SFS. It will be of great interest to use neutron scattering to look for the $2k_F$ peaks in the static spin structure factor $S(\vec{k})$. Furthermore, the SFS state is expected to have  low energy excitations concentrated around $2k_F$.   Gapless excitations are also expected for  $k<2k_F$ and to extend to an energy scale which is a fraction of the spinon bandwidth. This is seen in calculations based on the free spinon model~\cite{Li2017}.  The SFS is expected to have finite spin susceptibility. Experimental there is indeed a residual temperature independent susceptibility~\cite{Klanjvsek2017} but it is not known how much of it is spin or orbital in origin. We also expect a linear term $\gamma$ in the specific heat. This is seen experimentally but it is suppressed by a magnetic field~\cite{Ribak2017}. This contribution has been explained as mainly due to local moment~\cite{Kimchi2018}. If we use the large magnetic field limit to extract an intrinsic $\gamma$, we find a value of about $0.1 \text{mJ/K}^2$ per mole or $1.3 \text{mJ/K}^2$ per mole of star of David cluster. This is about a factor of 10 smaller than what is observed in the organics~\cite{Shimizu2003, Itou2007} , suggesting that the exchange energy scale is a factor of 10 larger. Finally the definitive evidence for a spinon Fermi surface will be the observation of a linear $T$ term in the thermal conductivity. While a null result was found earlier~\cite{Yu2017}, such a linear T term has recently been reported~\cite{murayama2018coexisting}. 

An exciting future avenue is the doping of the weak Mott insulator. Doping of the spinon Fermi surface spin liquid will likely lead to a correlation driven superconductor~\cite{Anderson1987}. As discussed earlier~\cite{Law2017}, the interesting regime is a doping concentration of about $10\%$ per cluster which is less than $1\%$ per Ta. Carrier localization is a serious challenge at such a  low doping level and gating of atomically thin samples may be the preferred method. The recent success of inducing superconductivity by doping the weak Mott insulator in twisted bilayer graphene is certainly encouraging~\cite{cao2018unconventional}. The 1T-TaS$_2$ system has the advantage of having a much higher temperature scale compared with graphene and the organics.

{\it Acknowledgments}\,---\, X.Y. Xu thanks the discussion with E.M. Stoudenmire and Donna Sheng. The calculations are performed using the ITensor C++ library (version 2.1.1). W.-Y. He, X.Y. Xu and K.T.L. thank the support of HKRGC through HKUST3/CRF/13G and C6026-16W. W.-Y. He also acknowledges the support of  Hong Kong PhD Fellowship.  G.C. is supported by the Ministry of Science and Technology of China with the Grant No.2016YFA0301001 and the Thousand-Youth-Talent Program of China.
 K.T.L. is further supported by the Croucher Foundation and the Dr Tai-chin Lo Foundation. P.A.L. acknowledges support by the US Department of Energy, Basic Energy Sciences under grant DE-FG02-03-ER46076.
He also thanks the hospitality of the Institute for Advanced Studies at the Hong Kong University of Science and Technology.
The simulation is performed at Tianhe-2 platform at the National Supercomputer Center in Guangzhou.

\bibliography{main}
\clearpage
\onecolumngrid
\begin{center}
\textbf{\large Supplemental Material for "Spinon Fermi surface in a cluster Mott insulator model on a triangular lattice and possible application to 1T-TaS$_2$"}
\end{center}
\setcounter{equation}{0}
\setcounter{figure}{0}
\setcounter{table}{0}
\setcounter{page}{1}
\makeatletter
\renewcommand{\thetable}{S\arabic{table}}
\renewcommand{\theequation}{S\arabic{equation}}
\renewcommand{\thefigure}{S\arabic{figure}}

\section{Effective tight binding Hamiltonian for the star of David superlattice}
\begin{figure}
\includegraphics[width=5in]{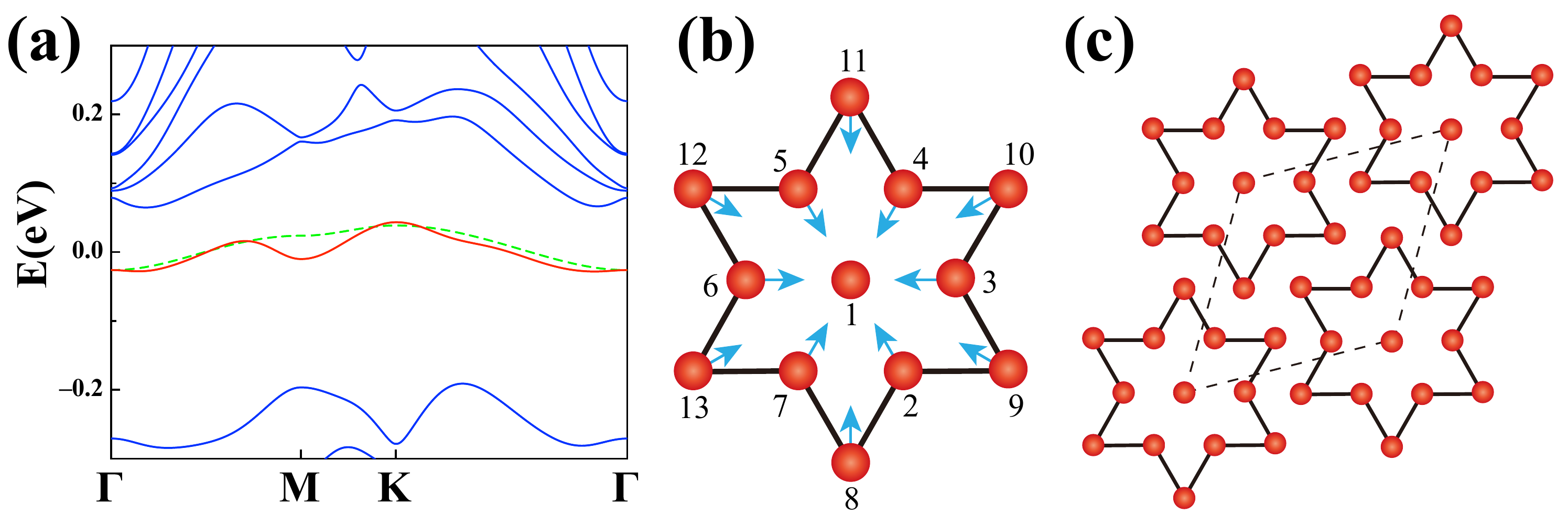}
\caption{(a) The band structure for the 1T-TaS$_2$ after reconstruction. The red band represents the isolated narrow band induced by the atomic SOC. The green dashed line is the energy dispersion from the effective two-orbital Hamiltonian. (b) The star of David unit cell. All the surrounding Ta atoms have displacement towards the centered Ta atom. (c) After the lattice deformation, the Ta atoms form the $\sqrt{13}\times\sqrt{13}$ star of David clusters and the clusters are further arranged as a triangular lattice.}
\label{FIGS1}
\end{figure}

In TaS$_2$, at the Fermi level, electrons are mainly from the 5d orbitals of the Ta atoms. Before lattice deformation, the $d_{z^2}$, $d_{xy}$ and $d_{x^2-y^2}$ orbitals from the Ta atoms are considered to construct the tight binding Hamiltonian as follows
\begin{align}
H_0=\sigma_0\otimes\begin{pmatrix}
f_0 & f_1 & f_2\\ 
f_1 & f_{11} & f_{12}\\ 
f_2 & f_{12} & f_{22}
\end{pmatrix}+\lambda_0\sigma_z\otimes\begin{pmatrix}
0 & 0 & 0\\ 
0 & 0 & i\\ 
0 & -i & 0
\end{pmatrix},
\end{align}
where the second term represents the atomic SOC. The hopping terms are
\begin{align}
f_0&=\epsilon_{z^2}+2t_0\left(2\cos\frac{1}{2}k_xa\cos\frac{\sqrt{3}}{2}k_ya+\cos k_xa\right),\\
f_{11}&=\epsilon_{xy}+2t_{11}\cos k_xa+\left(t_{11}+3t_{22}\right)\cos\frac{1}{2}k_xa\cos\frac{\sqrt{3}}{2}k_ya,\\
f_{22}&=\epsilon_{x^2-y^2}+2t_{22}\cos k_xa+\left(3t_{11}+2t_{22}\right)\cos\frac{1}{2}k_xa\cos\frac{\sqrt{3}}{2}k_ya,\\
f_{1}&=2\sqrt{3}t_2\sin\frac{1}{2}k_xa\sin\frac{\sqrt{3}}{2}k_ya\\
f_{2}&=2t_2\left(\cos\frac{1}{2}k_xa\cos\frac{\sqrt{3}}{2}k_ya-\cos k_xa\right)\\
f_{12}&=\sqrt{3}\left(t_{22}-t_{11}\right)\sin\frac{1}{2}k_xa\sin\frac{\sqrt{3}}{2}k_ya,
\end{align}
with $a$ the lattice constant for the triangular lattice before deformation. All the tight binding parameters can be found in Table.\ref{table1}~\cite{Smith1985, Rossnagel2006}. In the low temperature, the 1T-TaS$_2$ undergoes the commensurate charge density wave transition and form the triangular superlattice with $\sqrt{13}\times\sqrt{13}$ star of David unit cell, as is seen in Fig. \ref{FIGS1} (b). In the star of David unit cell, as the outer twelve Ta atoms have displacement toward the centered Ta atom, the interatomic distance from the inner surrounding Ta atoms (labeled by 2-7) and outer surrounding Ta atoms (labeled by 8-13) to the centered Ta atom 1 shrink by $6.4\%$ and $4.4\%$ respectively. Since the hopping integral is scaled as $R^{-5}$ with $R$ the Ta-Ta distance~\cite{Heine1967}, the atomic bonds within the star of David are therefore strengthened in the intra-cluster Hamiltonian $H_{\textrm{intra}}$ while other bonds are weakened in the inter-cluster Hamiltonian $H_{\textrm{inter}}$. As a result, after the lattice deformation, it is found in the reconstructed band structure that a narrow band is well isolated at the Fermi level, as is shown in Fig. \ref{FIGS1} (a).

We first regard a single star of David cluster as a super-atom and consider only the intra-cluster tight binding Hamiltonian $H_{\textrm{intra}}$. Through numerical diagonalization, at the energy of the narrow band the Kramers doublet $\Psi^{\uparrow}_{\alpha}$, $\Psi^{\downarrow}_{\beta}$ are found to be the Wannier function that compose the narrow band. The Wannier orbitals can be expressed in terms of the linear combination of atomic orbitals from the 13 Ta atoms with the coefficients $\alpha_{1}, \alpha_{2}, ...,\alpha_{7}$
\begin{align}\nonumber
\Psi^{\uparrow}_{\alpha}&=\alpha_1\psi^{1\uparrow}_{d_{x^2-y^2}+id_{xy}}+\alpha_2\left(\psi^{2\uparrow}_{d_{x^2-y^2}+id_{xy}}+\psi^{3\uparrow}_{d_{x^2-y^2}+id_{xy}}+\psi^{4\uparrow}_{d_{x^2-y^2}+id_{xy}}+\psi^{5\uparrow}_{d_{x^2-y^2}+id_{xy}}+\psi^{6\uparrow}_{d_{x^2-y^2}+id_{xy}}+\psi^{7\uparrow}_{d_{x^2-y^2}+id_{xy}}\right)\\\nonumber
&+\alpha_3\left(\psi^{2\uparrow}_{d_{x^2-y^2}-id_{xy}}+e^{i\frac{2\pi}{3}}\psi^{3\uparrow}_{d_{x^2-y^2}-id_{xy}}+e^{-i\frac{2\pi}{3}}\psi^{4\uparrow}_{d_{x^2-y^2}-id_{xy}}+\psi^{5\uparrow}_{d_{x^2-y^2}-id_{xy}}+e^{i\frac{2\pi}{3}}\psi^{6\uparrow}_{d_{x^2-y^2}-id_{xy}}+e^{-i\frac{2\pi}{3}}\psi^{7\uparrow}_{d_{x^2-y^2}-id_{xy}}\right)\\\nonumber
&+\alpha_4\left(\psi^{2\uparrow}_{d_{z^2}}+e^{-i\frac{2\pi}{3}}\psi^{3\uparrow}_{d_{z^2}}+e^{i\frac{2\pi}{3}}\psi^{4\uparrow}_{d_{z^2}}+\psi^{5\uparrow}_{d_{z^2}}+e^{-i\frac{2\pi}{3}}\psi^{6\uparrow}_{d_{z^2}}+e^{i\frac{2\pi}{3}}\psi^{7\uparrow}_{d_{z^2}}\right)\\\nonumber
&+\alpha_5\left(\psi^{8\uparrow}_{d_{x^2-y^2}+id_{xy}}+\psi^{9\uparrow}_{d_{x^2-y^2}+id_{xy}}+\psi^{10\uparrow}_{d_{x^2-y^2}+id_{xy}}+\psi^{11\uparrow}_{d_{x^2-y^2}+id_{xy}}+\psi^{12\uparrow}_{d_{x^2-y^2}+id_{xy}}+\psi^{13\uparrow}_{d_{x^2-y^2}+id_{xy}}\right)\\\nonumber
&+\alpha_6\left(\psi^{8\uparrow}_{d_{x^2-y^2}-id_{xy}}+e^{i\frac{2\pi}{3}}\psi^{9\uparrow}_{d_{x^2-y^2}-id_{xy}}+e^{-i\frac{2\pi}{3}}\psi^{10\uparrow}_{d_{x^2-y^2}-id_{xy}}+\psi^{11\uparrow}_{d_{x^2-y^2}-id_{xy}}+e^{i\frac{2\pi}{3}}\psi^{12\uparrow}_{d_{x^2-y^2}-id_{xy}}+e^{-i\frac{2\pi}{3}}\psi^{13\uparrow}_{d_{x^2-y^2}-id_{xy}}\right)\\
&+\alpha_7\left(\psi^{8\uparrow}_{d_{z^2}}+e^{-i\frac{2\pi}{3}}\psi^{9\uparrow}_{d_{z^2}}+e^{i\frac{2\pi}{3}}\psi^{10\uparrow}_{d_{z^2}}+\psi^{11\uparrow}_{d_{z^2}}+e^{-i\frac{2\pi}{3}}\psi^{12\uparrow}_{d_{z^2}}+e^{i\frac{2\pi}{3}}\psi^{13\uparrow}_{d_{z^2}}\right),
\end{align}
and $\Psi^{\downarrow}_{\beta}=i\sigma_y\Psi^{\uparrow\ast}_{\alpha}$ due to the time reversal symmetry, while the Wannier orbitals $\Psi^{\downarrow}_{\alpha}$, $\Psi^{\uparrow}_{\beta}$ are lifted in energy due to the atomic SOC. With the Wannier orbitals for the star of David super-atom, in the basis of $\left[\Psi^{\uparrow}_{\alpha}, \Psi^{\uparrow}_{\beta}, \Psi^{\downarrow}_{\alpha}, \Psi^{\downarrow}_{\beta}\right]^{\textrm{T}}$ the minimum two orbital Hamiltonian can be constructed as
\begin{align}\label{Eq1}
H=\begin{pmatrix}
2tC+\mu & 2t_1C_{-} & 0 & 0\\ 
2t_1^{\ast}C_{+} & 2tC+\mu & 0 & 0\\ 
0 & 0 & 2tC+\mu & 2t_1C_{-}\\ 
0 & 0 & 2t_1^{\ast}C_{+} & 2tC+\mu
\end{pmatrix}-\frac{1}{2}\lambda\sigma_z\otimes\tau_z,
\end{align}
with $t=-0.0072 \textrm{eV}, \  t_1=\left(-0.015+0.0068i\right)\textrm{eV}, \ \mu=0.2897 \textrm{eV}, \  \lambda=0.5622 \textrm{eV}$, and $\sigma$, $\tau$ for the spin, orbital space respectively. The matrix element for $H$ is obtained through $H_{i, j}=\bra{\Psi_{i}}H_{\textrm{intra}}+H_{\textrm{inter}}\ket{\Psi_{j}}$. The energy dispersion for the narrow band from the minimum two orbital model is plotted as the green dashed line in Fig.\ref{FIGS1} for comparison. Here the basis functions in Eq. \ref{Eq1} read
\begin{align}\nonumber
C&=\cos\left(\frac{7}{2}k_x+\frac{\sqrt{3}}{2}k_y\right)a+\cos\left(-\frac{5}{2}k_x+\frac{3\sqrt{3}}{2}k_y\right)a+\cos\left(k_x-2\sqrt{3}k_y\right)a\\\nonumber
C_+&=\cos\left(\frac{7}{2}k_x+\frac{\sqrt{3}}{2}k_y\right)a+e^{i\frac{2\pi}{3}}\cos\left(-\frac{5}{2}k_x+\frac{3\sqrt{3}}{2}k_y\right)a+e^{i\frac{4\pi}{3}}\cos\left(k_x-2\sqrt{3}k_y\right)a\\
C_-&=\cos\left(\frac{7}{2}k_x+\frac{\sqrt{3}}{2}k_y\right)a+e^{-i\frac{2\pi}{3}}\cos\left(-\frac{5}{2}k_x+\frac{3\sqrt{3}}{2}k_y\right)a+e^{-i\frac{4\pi}{3}}\cos\left(k_x-2\sqrt{3}k_y\right)a.
\end{align}

\begin{table}
\centering
\caption{The tight binding parameters in the unit of eV.} 
\begin{tabular}{c c c c c c c c c c c} 
\hline\hline 
$\epsilon_{z^2}$ & $\epsilon_{xy}$ & $\epsilon_{x^2-y^2}$ & $t_0$ & $t_{11}$ & $t_{22}$ & $t_2$ & $\lambda_0$\\ [0.5ex] 
\hline 
1.4052 & 1.3440 & 1.3440 & -0.1046 & 0.2406 & -0.5320 & -0.3701 & -0.3130\\ [1ex] 
\hline 
\end{tabular}
\label{table1} 
\end{table}

\section{XXZ model from the two-orbital Hubbard model}
\begin{figure}
\includegraphics[width=6in]{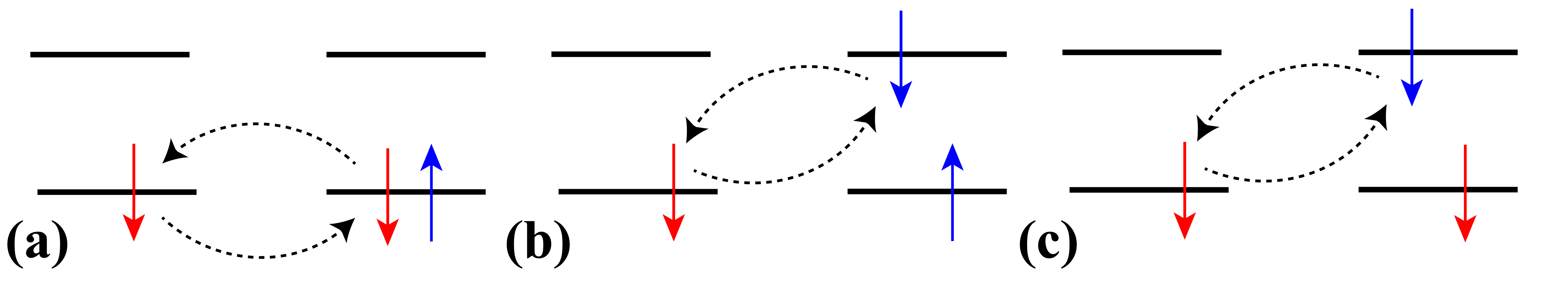}
\caption{The spin exchange between two sites involving the two orbitals. The blue arrows represent electrons from $\alpha$ orbital while the red arrows represent electrons from $\beta$ orbital. In (a), only the intra-orbital hopping is involved while in (b) and (c), only the inter-orbital hopping is involved. For the intermediate excited states, the excitation energy in (a) and (c) is from the inter-orbital interaction while in (b) it is intra-orbital interaction.}
\label{FIGS2}
\end{figure}

With the two Wannier orbitals for the star of David super-atom, considering both the intra-orbital interaction $U$ and inter-orbital interaction $U_1$, we can construct the two-orbital Hubbard model as
\begin{align}
\mathcal{H}=\mathcal{H}_0+\mathcal{H}_1,
\end{align}
with
\begin{align}\nonumber
\mathcal{H}_0&=\sum_i\left({U_0c^{\dagger}_{i\alpha\uparrow}c_{i\alpha\uparrow}c^{\dagger}_{i\alpha\downarrow}c_{i\alpha\downarrow}+U_0c^{\dagger}_{i\beta\uparrow}c_{i\beta\uparrow}c^{\dagger}_{i\beta\downarrow}c_{i\beta\downarrow}+U_1c^{\dagger}_{i\alpha\uparrow}c_{i\alpha\uparrow}c^{\dagger}_{i\beta\downarrow}c_{i\beta\downarrow}+U_1c^{\dagger}_{i\alpha\downarrow}c_{i\alpha\downarrow}c^{\dagger}_{i\beta\uparrow}c_{i\beta\uparrow}+U_1c^{\dagger}_{i\alpha\uparrow}c_{i\alpha\uparrow}c^{\dagger}_{i\beta\uparrow}c_{i\beta\uparrow}}\right.\\
&\left.{+U_1c^{\dagger}_{i\alpha\downarrow}c_{i\alpha\downarrow}c^{\dagger}_{i\beta\downarrow}c_{i\beta\downarrow}}\right)+\sum_i\lambda\left(c^{\dagger}_{i\beta\uparrow}c_{i\beta\uparrow}+c^{\dagger}_{i\alpha\downarrow}c_{i\alpha\downarrow}\right),
\end{align}
and
\begin{align}
\mathcal{H}_1=\sum_{\left \langle i, j \right \rangle,\sigma}\left(tc^{\dagger}_{i\alpha\sigma}c_{j\alpha\sigma}+tc^{\dagger}_{i\beta\sigma}c_{j\beta\sigma}+t_1c^{\dagger}_{i\alpha\sigma}c_{j\beta\sigma}+t_1^{\ast}c^{\dagger}_{i\beta\sigma}c_{j\alpha\sigma}\right)+h.c.,
\end{align}
where $c^{\left(\dagger\right)}_{i,\alpha\left(\beta\right)\sigma}$ is the annihilation (creation) operator that annihilates (creates) an $\alpha$ ($\beta$) Wannier orbital state with spin state $\sigma$ at the site $i$. In order to get the effective spin model, we first analyze the neighboring star of David clusters at the site $i$, $j$. At quarter filling for the two-orbital Hubbard model, the states with single occupancy, listed as the first row in the table.\ref{Basis}, form the subspace with low energy. Through kinetic hopping, these states can transfer to the excited states listed in the remaining part of table.\ref{Basis}. As a result, writing the two-orbital Hubbard model in the basis from table.\ref{Basis}, we can perturbatively deal with the excited states through the second order Schrieffer-Wolff transformation and obtain the effective Hamiltonian in the low energy subspace
\begin{align}
\mathcal{H}^{\left(2\right)}=\frac{4t^2}{U_1}\left(S^x_iS^y_j+S^y_iS^y_j\right)+4\left(\frac{t^2}{U_1}+\frac{t_1t^{\ast}_1}{U_0+\lambda}-\frac{t_1t^{\ast}_1}{U_1+\lambda}\right)S^z_iS^z_j-4\left(\frac{t^2}{U_1}+\frac{t_1t^{\ast}_1}{U_0+\lambda}+\frac{t_1t^{\ast}_1}{U_1+\lambda}\right),
\end{align}
which is the XXZ model. For the neighboring star of David clusters, the spin exchange mechanism is schematically shown in Fig. \ref{FIGS2}. As the two Wannier orbitals $\Psi_{\alpha}$ and $\Psi_{\beta}$ are both involved due to the atomic SOC, the spin exchange can go through either the intra-orbital hopping or the inter-orbital hopping. In the process in Fig. S2 (a), the exchange through the intra-orbital hopping gives rise to the Heisenberg model as usual. In the process in Fig. S2 (b) and (c), the virtual hopping between the low energy state and the excited state is inter-orbital type and the low energy state cannot be changed, so the inter-orbital type virtual hopping results in the $S^z$ terms in the spin model. Since the intermediate states in the process (b) and (c) have intra-orbital interaction and inter-orbital interaction as the excitation energy respectively, the two process cannot be canceled out and eventually generate the anisotropy in the $S^z$ term. As a result, considering the neighboring star of David clusters we obtain the XXZ spin model.

\begin{table}
\centering
\caption{The basis for the neighboring star of David clusters}
\begin{tabular}{c|cccc} \hline
$n$ & \multicolumn{4}{c}{$\Psi_n\in\mathcal{H}$} \\
\hline
$ 1\sim4 $ & $ \ket{i\alpha\uparrow,j\alpha\uparrow} $ &  $ \ket{i\beta\downarrow,j\alpha\uparrow} $  & $ \ket{i\alpha\uparrow,j\beta\downarrow} $ & $ \ket{i\beta\downarrow,j\beta\downarrow} $\\\hline
$ 5\sim8 $ & $ \ket{i\alpha\uparrow\beta\downarrow,0} $ & $ \ket{0,j\alpha\uparrow\beta\downarrow} $ & $ \ket{0,j\alpha\uparrow\alpha\downarrow} $ & $ \ket{i\alpha\uparrow\alpha\downarrow,0} $\\
\hline
$ 9\sim12 $ & $ \ket{i\beta\uparrow\beta\downarrow,0} $ & $ \ket{0,j\beta\uparrow\beta\downarrow} $ & $ \ket{0,j\alpha\uparrow\beta\uparrow} $ & $ \ket{i\alpha\uparrow\beta\uparrow,0} $\\
\hline
$ 13\sim14 $ & $ \ket{i\alpha\downarrow\beta\downarrow,0} $ & $ \ket{0,j\alpha\downarrow\beta\downarrow} $ & &  \\
\hline
\end{tabular}
\label{Basis}
\end{table}

\section{Extrapolation to the anisotropy modified ring exchange term}
\begin{figure}
\includegraphics[width=3.6in]{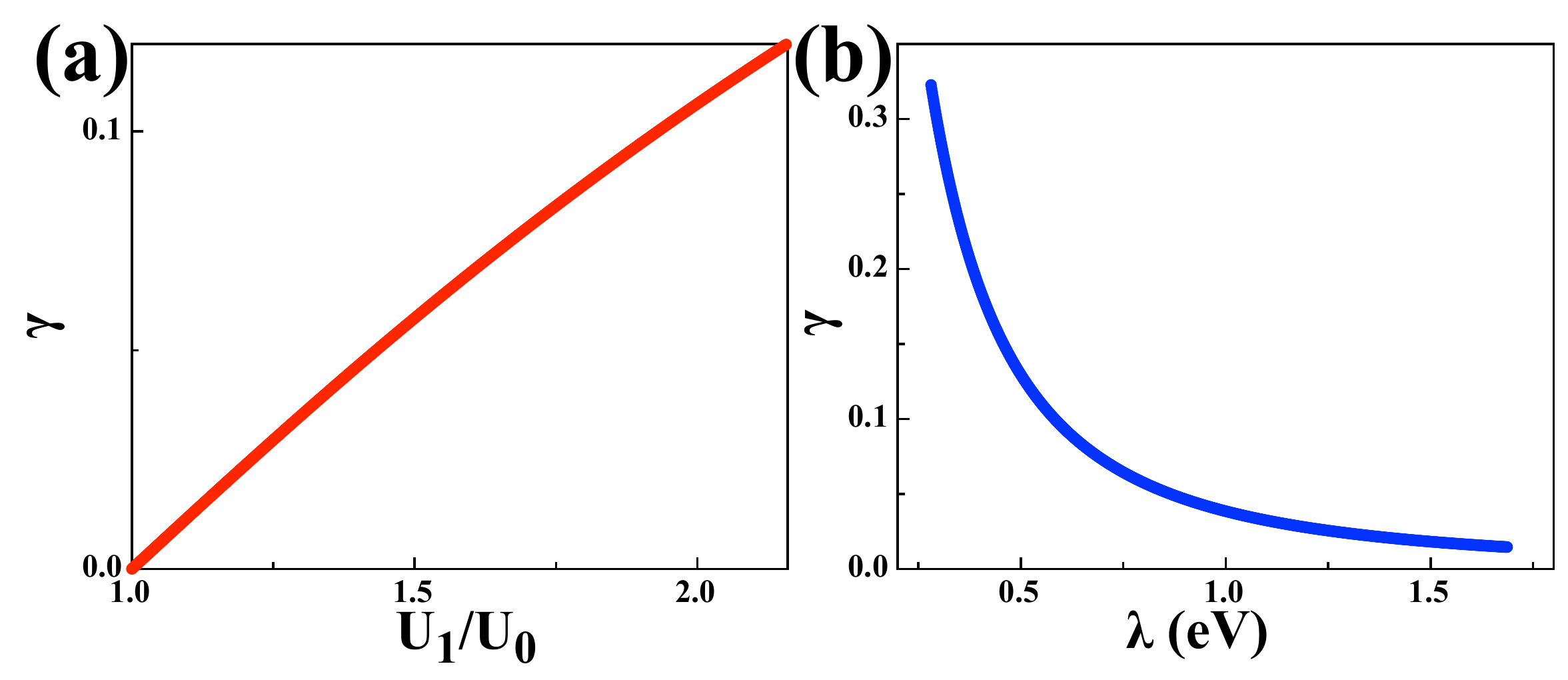}
\caption{(a) The anisotropy $\gamma$ increases with the increase of the ratio between the inter-orbital interaction $U_1$ and intra-orbital interaction $U_0$, the atomic SOC is taken to be $\lambda=0.5622\textrm{eV}$. (b) For fixed $\frac{U_1}{U_0}=2$, the anisotropy $\gamma$ decreases to 0 with the increase of the atomic SOC $\lambda$. In both the calculations, it is fixed to the Mott gap that $U_1+U_0=200\textrm{meV}$.}
\label{FIGS3}
\end{figure}

In the neighboring two clusters analysis, as the anisotropic term in the XXZ model arises from the interplay of inter-orbital hopping, inter-orbital interaction and intra-orbital interaction, we would like to explore the possibility to reduce the two-orbital complexity to an effective one orbital Hubbard model with pseudo-spin states and simplify the whole system. We notice that including both the low energy states and excited states, the effective Hamiltonian $\mathcal{H}^{\left(2\right)}$ from the second order Schrieffer-Wolff transformation in the basis of $c^{\dagger}_{j\alpha\uparrow}c^{\dagger}_{i\beta\downarrow}\ket{0}$, $c^{\dagger}_{j\beta\downarrow}c^{\dagger}_{i\alpha\uparrow}\ket{0}$, $c^{\dagger}_{i\alpha\uparrow}c^{\dagger}_{i\beta\downarrow}\ket{0}$, $c^{\dagger}_{j\alpha\uparrow}c^{\dagger}_{j\beta\downarrow}\ket{0}$ reads
\begin{align}
\mathcal{H}^{\left(2\right)}=\begin{pmatrix}
-\frac{2t^2}{U_1}+\frac{2t_1t_1^{\ast}}{U_1+\lambda}-\frac{2t_1t_1^{\ast}}{U_0+\lambda} & \frac{2t^2}{U_1} & 0 & 0\\ 
\frac{2t^2}{U_1} & -\frac{2t^2}{U_1}+\frac{2t_1t_1^{\ast}}{U_1+\lambda}-\frac{2t_1t_1^{\ast}}{U_0+\lambda} & 0 & 0\\ 
0 & 0 & \frac{2t_1t_1^{\ast}}{U_1+\lambda}+\frac{2t^2}{U_1}+U_1 & \frac{2t^2}{U_1}\\ 
0 & 0 & \frac{2t^2}{U_1} & \frac{2t_1t_1^{\ast}}{U_1+\lambda}+\frac{2t^2}{U_1}+U_1
\end{pmatrix}.
\end{align}
We take $c^{\dagger}_{\alpha\uparrow}\ket{0}$ and $c^{\dagger}_{\beta\downarrow}\ket{0}$ as the pseudo-spin states and construct an effective one orbital Hubbard model $\tilde{\mathcal{H}}$ so that under Schrieffer-Wolff transformation the effective Hamiltonian $\mathcal{H}^{\left(2\right)}$ is recovered
\begin{align}\nonumber
\mathcal{H}^{\left(2\right)}&=e^{S}\tilde{\mathcal{H}}e^{-S}\\
&\approx\tilde{\mathcal{H}}+\left[S, \tilde{\mathcal{H}}\right]+\frac{1}{2}\left[S, \left[S, \tilde{\mathcal{H}}_{\textrm{diag}}\right]\right],
\end{align}
where $\tilde{\mathcal{H}}_{\textrm{diag}}$ is the diagonal part of the Hamiltonian matrix $\tilde{\mathcal{H}}$ and
\begin{align}
\tilde{\mathcal{H}}=\begin{pmatrix}
0 & 0 & t_\alpha & t_\beta\\ 
0 & 0 & -t_\beta & -t_\alpha\\ 
t_\alpha & -t_\beta & U & 0\\ 
t_\beta & -t_\alpha & 0 & U
\end{pmatrix}, \quad S=\frac{1}{U}\begin{pmatrix}
0 & 0 & -t_\alpha & -t_\beta\\ 
0 & 0 & t_\beta & t_\alpha\\ 
t_\alpha & -t_\beta & 0 & 0\\ 
t_\beta & -t_\alpha & 0 & 0
\end{pmatrix},
\end{align}
with the modified on-site interaction and kinetic hopping
\begin{align}
U&=\frac{4t_1t^{\ast}_1}{U_1+\lambda}+U_1-\frac{2t_1t^{\ast}_1}{U_0+\lambda},\\
t_{\alpha}&=\frac{1}{2}\left[\sqrt{\left(\frac{4t^2}{U_1}+\frac{2t_1t^{\ast}_1}{U_0+\lambda}-\frac{2t_1t^{\ast}_1}{U_1+\lambda}\right)\left(\frac{4t_1t_1^{\ast}}{U_1+\lambda}+U_1-\frac{2t_1t^{\ast}_1}{U_0+\lambda}\right)}+\sqrt{\left(\frac{2t_1t^{\ast}_1}{U_0+\lambda}-\frac{2t_1t^{\ast}_1}{U_1+\lambda}\right)\left(\frac{4t_1t^{\ast}_1}{U_1+\lambda}+U_1-\frac{2t_1t^{\ast}_1}{U_0+\lambda}\right)}\right],\\
t_{\beta}&=\frac{1}{2}\left[\sqrt{\left(\frac{4t^2}{U_1}+\frac{2t_1t^{\ast}_1}{U_0+\lambda}-\frac{2t_1t^{\ast}_1}{U_1+\lambda}\right)\left(\frac{4t_1t_1^{\ast}}{U_1+\lambda}+U_1-\frac{2t_1t^{\ast}_1}{U_0+\lambda}\right)}-\sqrt{\left(\frac{2t_1t^{\ast}_1}{U_0+\lambda}-\frac{2t_1t^{\ast}_1}{U_1+\lambda}\right)\left(\frac{4t_1t^{\ast}_1}{U_1+\lambda}+U_1-\frac{2t_1t^{\ast}_1}{U_0+\lambda}\right)}\right].
\end{align}
We write the effective one orbital Hubbard model $\tilde{\mathcal{H}}$ in the form of annihilation and creation operator
\begin{align}
\tilde{\mathcal{H}}=\sum_{\left \langle i, j \right \rangle}\left(t_{\alpha}\tilde{c}^{\dagger}_{i\alpha\uparrow}\tilde{c}_{j\alpha\uparrow}+t_{\beta}\tilde{c}^{\dagger}_{i\beta\downarrow}\tilde{c}_{j\beta\downarrow}\right)+\sum_{i}U\tilde{c}^{\dagger}_{i\alpha\uparrow}\tilde{c}_{i\alpha\uparrow}\tilde{c}^{\dagger}_{i\beta\downarrow}\tilde{c}_{i\beta\downarrow}+h.c.,
\end{align}
and then apply the Schrieffer-Wolff transformation to generate the effective spin model~\cite{Macdonald1988}. For the neighboring two star of David clusters, we recover the XXZ model $\tilde{\mathcal{H}}^{\left(2\right)}=\mathcal{H}^{\left(2\right)}$. For a diamond cluster of star of Davids, we get the anisotropy modified ring exchange term
\begin{align}\nonumber\label{Eq2}
\tilde{\mathcal{H}}^{\left(4\right)}&=\frac{80}{U^3}\left\{{\left[t_{\alpha}t_{\beta}\left(S_i^xS_j^x+S_i^yS_j^y\right)+\frac{t_{\alpha}^2+t_{\beta}^2}{2}S_i^zS_j^z\right]\left[t_{\alpha}t_{\beta}\left(S_k^xS_l^x+S_k^yS_l^y\right)+\frac{t_{\alpha}^2+t_{\beta}^2}{2}S_k^zS_l^z\right]}\right.\\
&\left.{+\left[t_{\alpha}t_{\beta}\left(S_j^xS_k^x+S_j^yS_k^y\right)+\frac{t_{\alpha}^2+t_{\beta}^2}{2}S_j^zS_k^z\right]\left[t_{\alpha}t_{\beta}\left(S_i^xS_l^x+S_i^yS_l^y\right)+\frac{t_{\alpha}^2+t_{\beta}^2}{2}S_i^zS_l^z\right]-t_{\alpha}^2t_{\beta}^2\left({\bm S}_i\cdot{\bm S}_k\right)\left({\bm S}_j\cdot{\bm S}_l\right)}\right\}.
\end{align}
Eventually the general form for the effective spin model of the 1T-TaS$_2$ is derived as
\begin{align}\nonumber
\tilde{\mathcal{H}}_{\textrm{eff}}&=J\sum_{\left \langle i,j \right \rangle}\left(S_i^xS_j^x+S_i^yS_j^y+(1+\gamma) S_i^zS_j^z\right)+K\sum_{\left \langle i, j, k, l \right \rangle}\left[{\left(S_i^xS_j^x+S_i^yS_j^y+(1+\gamma) S_i^zS_j^z\right)\left(S_k^xS_l^x+S_k^yS_l^y+(1+\gamma) S_k^zS_l^z\right)}\right.\\
&\left.{+\left(S_j^xS_k^x+S_j^yS_k^y+(1+\gamma) S_j^zS_k^z\right)\left(S_i^xS_l^x+S_i^yS_l^y+(1+\gamma) S_i^zS_l^z\right)-\left({\bm S}_i\cdot{\bm S}_k\right)\left({\bm S}_j\cdot{\bm S}_l\right)}\right].
\label{Heffsupp}
\end{align}
Here $\gamma=\frac{(t_\alpha-t_\beta)^2}{2t_\alpha t_\beta}$ denotes the anisotropy of the spin model, and the anisotropy $\gamma$ can be estimated with Eq. S5 and Eq. S6 as is shown in Fig.~\ref{FIGS3} (a). In the estimation, we fix the $U_1+U_0=200\textrm{meV}$, which is about the Mott gap measured in the tunneling spectroscopy~\cite{Qiao2017}, and vary the ratio between the inter-orbital interaction $U_1$ and the intra-orbital interaction $U_0$ to obtain the evolution of the anisotropy $\gamma$. It can be seen that the anisotropy $\gamma$ increases slowly with the increase of $\frac{U_1}{U_0}$ and $\gamma$ reaches 0.1 when the $U_1=2U_0$. If we further increase the ratio $\frac{U_1}{U_0}\rightarrow\infty$, the anisotropy will reach a constant value around 0.4 and converge. In the limit of $\frac{U_1}{U_0}\rightarrow\infty$, the intra-orbital interaction is eliminated and the anisotropy is only affected by the atomic SOC $\lambda$. In the two orbital Hubbard model, as the gap between the two bands increases with the increase of atomic SOC, it regress to the single orbital Hubbard model and the anisotropy $\gamma$ will approach $0$ in large atomic SOC, as is shown in Fig.~\ref{FIGS3} (b).

\section{DMRG simulation setup}
\begin{figure}[b]
\includegraphics[width=4in]{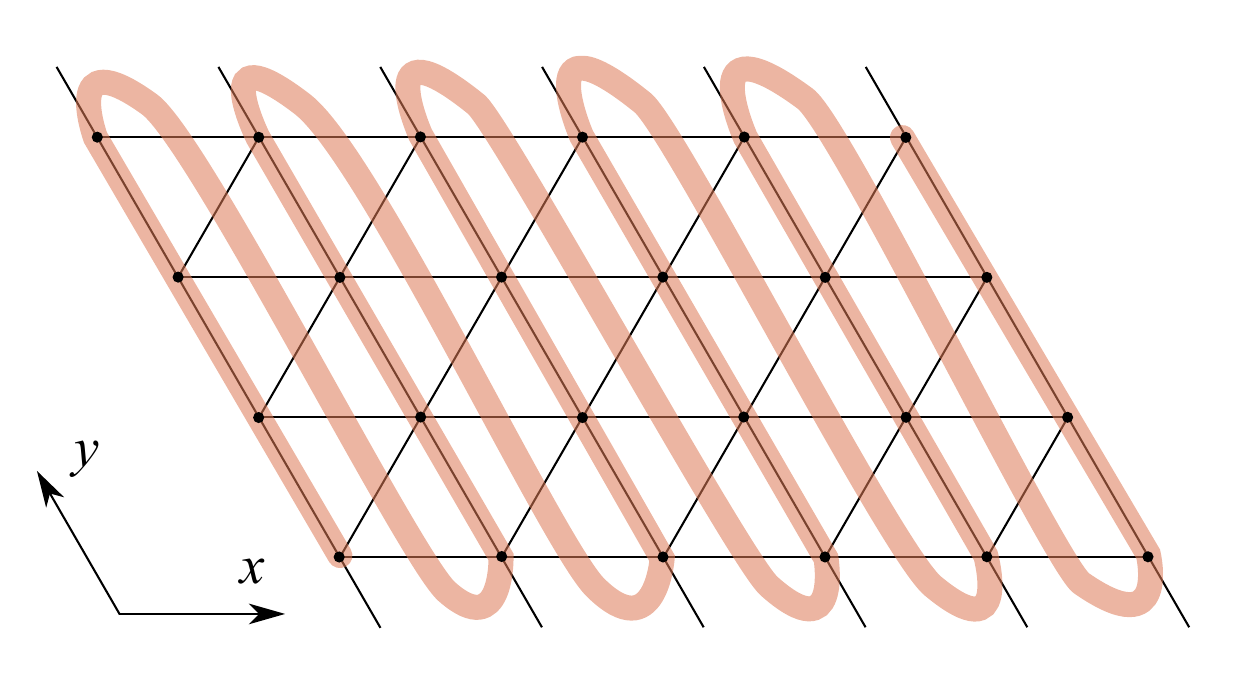}
\caption{Schematic DMRG path (showed with shadowed red line) for a ($N_y=4, N_x=6$) cluster. The black line denotes the nearest neighbor two spin exchange bonds, the four-spin exchange terms are not showed in the figure. In the simulation, $y$-direction periodic condition is used, while $x$-direction boundary is open.}
\label{dmrgpath}
\end{figure}

As we already discussed in the main text, several limits of effective spin Hamiltonian~\eqref{Heffsupp} have been widely explored, but for general parameter region, especially in the situation strongly related to real materials 1T-TaS$_2$ where $\gamma \lesssim 0.1$ and with a meaningful large of $K/J$, the physics is still unknown.  To identify possible ground state around this region, we use DMRG to simulate the effective spin Hamiltonian~\eqref{Heffsupp}.
The  MPS representation is used  in our DMRG simulation and calculations are performed using the ITensor C++ library (version 2.1.1), http://itensor.org/. The model has total $S^z$ conservation, and all results are got in $S_{\text{tot}}^z=0$ sector if it is not specified. We use the cylinder geometry with open boundary condition in $x$ direction, as showed in Fig.~\ref{dmrgpath}.

\section{Determine phase diagram}
\subsection{Identify different phases with various correlation functions}
We will mainly focus on $\gamma=0$ as in 1T-TaS$_2$, anisotropy $\gamma$ is small enough to be neglected. To identify different phases, various correlation functions are measured, including spin correlation, dimer correlation
and chirality correlation.

Spin correlation $C_{S}(i,i')=\langle\vec{S}_{i}\cdot\vec{S}_{i'}\rangle$ with
\begin{equation}
\vec{S}_{i}\cdot\vec{S}_{i'}=\frac{1}{2}\left(S_{i}^{+}S_{i'}^{-}+S_{i}^{-}S_{i'}^{+}\right)+S_{i}^{z}S_{i'}^{z}
\label{eq:spincorr}
\end{equation}

Dimer correlation $C_{D_b}(i,i')=\langle D_{b}(i)D_{b}(i')\rangle-\langle D_b\rangle^{2}$,
where the dimer $D_{b}=\vec{S}_{i}\cdot\vec{S}_{i+\delta}$ can have three types labeled by $b=x,y,xy$ due to the three possible directions
of bonds, namely $\delta=\hat{x}$, $\delta=\hat{y}$ or $\delta=\hat{x}+\hat{y}$.

\begin{equation}
D_{b}(i)D_{b}(i')=\left(\frac{1}{2}\left(S_{i}^{+}S_{i+\delta}^{-}+S_{i}^{-}S_{i+\delta}^{+}\right)+S_{i}^{z}S_{i+\delta}^{z}\right)\left(\frac{1}{2}\left(S_{i'}^{+}S_{i'+\delta}^{-}+S_{i'}^{-}S_{i'+\delta}^{+}\right)+S_{i'}^{z}S_{i'+\delta}^{z}\right)
\end{equation}

Chirality correlation $C_{X}(\triangle,\triangle')=\langle X_{\triangle}X_{\triangle'}\rangle$,
where the spin chirality $X_{\triangle}$ is defined as $X_{\triangle}=\vec{S}_{i}\cdot\left(\vec{S}_{j}\times\vec{S}_{k}\right)$
(with $i,j,k\in\triangle$).

\begin{equation}
X_{\triangle}=\frac{i}{2}\left(S_{i}^{+}S_{j}^{-}-S_{i}^{-}S_{j}^{+}\right)S_{k}^{z}+\frac{i}{2}\left(S_{k}^{+}S_{i}^{-}-S_{k}^{-}S_{i}^{+}\right)S_{j}^{z}+\frac{i}{2}\left(S_{j}^{+}S_{k}^{-}-S_{j}^{-}S_{k}^{+}\right)S_{i}^{z}
\end{equation}
\begin{eqnarray}
X_{\triangle}X_{\triangle'} & = & -\frac{1}{4}\left[\left(S_{i}^{+}S_{j}^{-}-S_{i}^{-}S_{j}^{+}\right)S_{k}^{z}+\left(S_{k}^{+}S_{i}^{-}-S_{k}^{-}S_{i}^{+}\right)S_{j}^{z}+\left(S_{j}^{+}S_{k}^{-}-S_{j}^{-}S_{k}^{+}\right)S_{i}^{z}\right]\nonumber \\
 &  & \left[\left(S_{i'}^{+}S_{j'}^{-}-S_{i'}^{-}S_{j'}^{+}\right)S_{k'}^{z}+\left(S_{k'}^{+}S_{i'}^{-}-S_{k'}^{-}S_{i'}^{+}\right)S_{j'}^{z}+\left(S_{j'}^{+}S_{k'}^{-}-S_{j'}^{-}S_{k'}^{+}\right)S_{i'}^{z}\right]
\end{eqnarray}

After the DMRG energy optimization, we will get the ground state wavefunction. Measuring above correlations is tedious and slow for large system size and  large bond dimension. We need to take advantage of the mixed-canonical matrix product state (matrices to the left are left-normalized, to the right are
right-normalized.) and classify the correlations with different site indexes into several classes and measure them separately. With this trick, the measurement is highly accelerated and thus we are able to measure all the correlation mentioned above with limited computation resources we have.

When we get the real space correlations, a fast way to check whether there is any order is to first perform the Fourier transformation to the momentum space and get the structure factor, and then check whether there is any peak in the structure factor.
This first step usually will help us identify whether there exists certain ordered phase, like spin order, bond order, etc. If we do not see any peak (ignore finite size broadening), then we will roughly rule out any orders which result in peaks in the corresponding structure factor and it will be more like a QSL phase, although more evidences are needed.

\renewcommand{\arraystretch}{1.5}
\begin{table}[ht]
\caption{The necessary conditions for detecting various spin liquid (SL) phase in DMRG simulation.} 
\centering 
\begin{tabular}{c | c | c | c | c} 
\hline\hline
\multicolumn{2}{c|} {\textbf{type of SL}}  & \textbf{structure factor}  & \textbf{real space decay} & \textbf{static spin susceptibility}\\
\hline
\multirow{3}{*}{gapless SL} & spinon Fermi surface & $2k_F$ peaks & power law & finite \\\cline{2-5}
 & Z2 Dirac SL & - & power law & linear in $T$ \\\cline{2-5}
 & U(1) Dirac SL & - & universal power law & linear in $T$\\
\hline
\multirow{3}{*}{gapped SL} & Z2 SL & - & exponential & exponentially small \\\cline{2-5}
 & U(1) SL\textsuperscript{*}& - & exponential & exponentially small \\\cline{2-5}
 & chiral SL & chiral order & exponential & exponentially small \\
\hline
\multicolumn{5}{l}{\textsuperscript{*}\footnotesize{not allowed in 2D}}
\end{tabular}
\label{table:conditionforSL} 
\end{table}

One easily obtainable evidence is the functional form of the real space decay of different correlations. The real space decay of the spin correlation function will reflect the spin gap information, a short correlation length is consistent with a spin gap, while a decay of universal power law $\sim r^{-4}$ is consistent with (algebraic) U(1) Dirac spin liquid which is gapless. Actually, for algebraic spin liquid as a critical phase, the dimer correlation will also follow the same power law decay. For other correlations, we can also get information from the real space decay. For example, the short correlation length of dimer correlations is consistent with a singlet gap, while the long range chiral correlation is consistent with time-reversal broken chiral spin liquid phase.
In addition, detailed analysis of momentum space structure will also provide further evidences. As discussed in Refs~\cite{Sheng2009,Block2011}, a detailed analysis of the $2k_F$ peaks in structure factor should help to identify the gapless QSL with spinon Fermi surface.

\subsection{Distinguish gapped and gapless phase with static spin susceptibility}
Another way to distinguish gapped and gapless phase is to calculate static spin susceptibility. Here we can apply small magnetic fields along $z$-direction and measure the magnetic moment density $M$. As the $z$-direction magnetic field will not break the total $S^z$ conservation, the changing of magnetic field does not make the common eigenstates of field term and zero-field Hamiltonian evolve~\cite{Savary2017}, so the static spin susceptibility is directly related to the density of excitations. The finite static spin susceptibility will correspond to a gapless phase while a exponentially small static spin susceptibility will correspond to a gapped phase. For gapless Dirac phase, the static spin susceptibility will obey a linear $T$ dependence and will not be finite at zero temperature.

 In Table~\ref{table:conditionforSL}, we list almost all possible types of spin liquids and their corresponding necessary conditions which can be checked in DMRG simulation as we discussed in above.

\section{phase diagram}
The phase diagram is already showed in Fig.~2 of main text. 

\subsection{AFM phase}
\begin{figure}[h!]
\includegraphics[width=0.3\columnwidth]{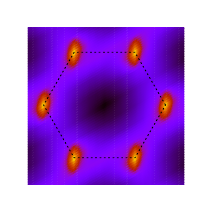}
\caption{Spin structure factor for $L_x=12$, $L_y=6$ at $K/J=0.08$ and $\gamma=0$, where we have $120^\circ$-spin order, with sharp peaks at $K$ points in spin structure factor. In above figure, dash line denotes Brillouin zone of triangular lattice.}
\label{fig:Sk_K008}
\end{figure}

\begin{figure}[h!]
\includegraphics[width=0.5\columnwidth]{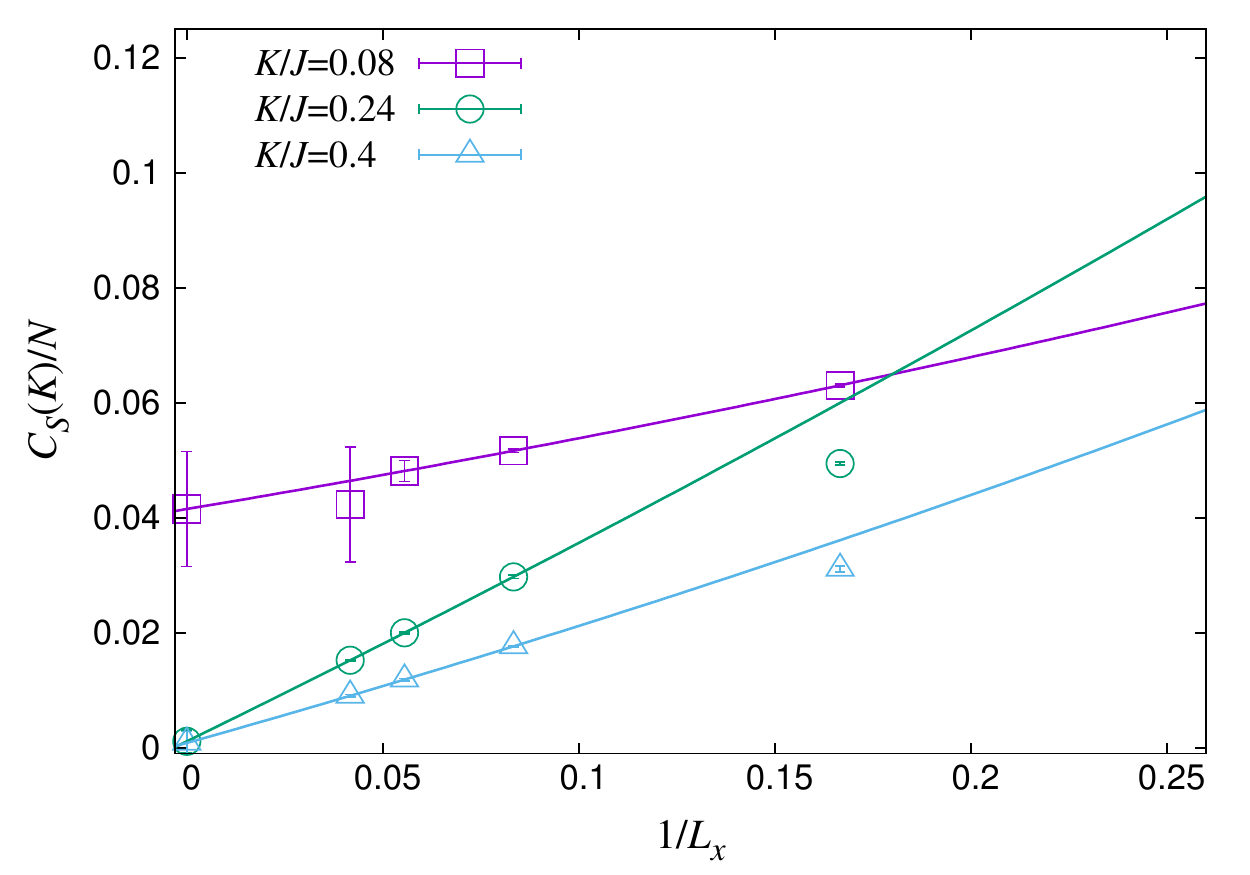}
\caption{The extrapolation of square of $120^\circ$-spin order ($C_{S}(K)/N$) to infinite $L_x$ for $L_y=6$, where $N=L_xL_y$. Obviously, at small $K/J$ region (here we plot a representative point $K/J=0.08$), there is prominent $120^\circ$-spin order.}
\label{fig:SextravsLx}
\end{figure}

The AFM phase has the in-plane $120^\circ$-spin order with $Q$ vector at the corner $K$ point ( $(\frac 1 3, \frac 1 3)$ in the basis of reciprocal lattice primitive vectors). Fig.~\ref{fig:Sk_K008} shows spin structure factor at $K/J=0.08$ with $L_x=12$ and $L_y=6$. The finite size scaling of this spin order is showed in Fig.~\ref{fig:SextravsLx} only for $L_y=6$ systems, while for 2D finite size scaling with fixing $L_x$ and $L_y$ ratio is not possible, as we need both $L_x$ and $L_y$ to be multiples of 6 and to have three points in $y$-direction $L_y=6$, $12$ and $18$ is out of current limit of computation power. The $120^\circ$-spin order is prominent in small $K/J$ region (for example $K/J=0.08$ in the Fig.~\ref{fig:SextravsLx}), and absent in larger $K/J$ region (for example $K/J=0.24$ and $0.40$ in the Fig.~\ref{fig:SextravsLx}).

\subsection{VBS phase}
\begin{figure}[h!]
\includegraphics[width=0.3\columnwidth]{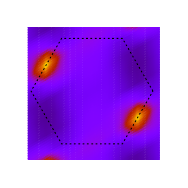}
\caption{$y$-direction dimer structure factor for $L_x=12$, $L_y=6$ at $K/J=0.24$ and $\gamma=0$, where we have dimer order, with sharp peaks at $M$ points in dimer structure factor. In above figure, dash line denotes Brillouin zone of triangular lattice.}
\label{fig:Dyk_M024}
\end{figure}

\begin{figure}[h!]
\includegraphics[width=0.5\columnwidth]{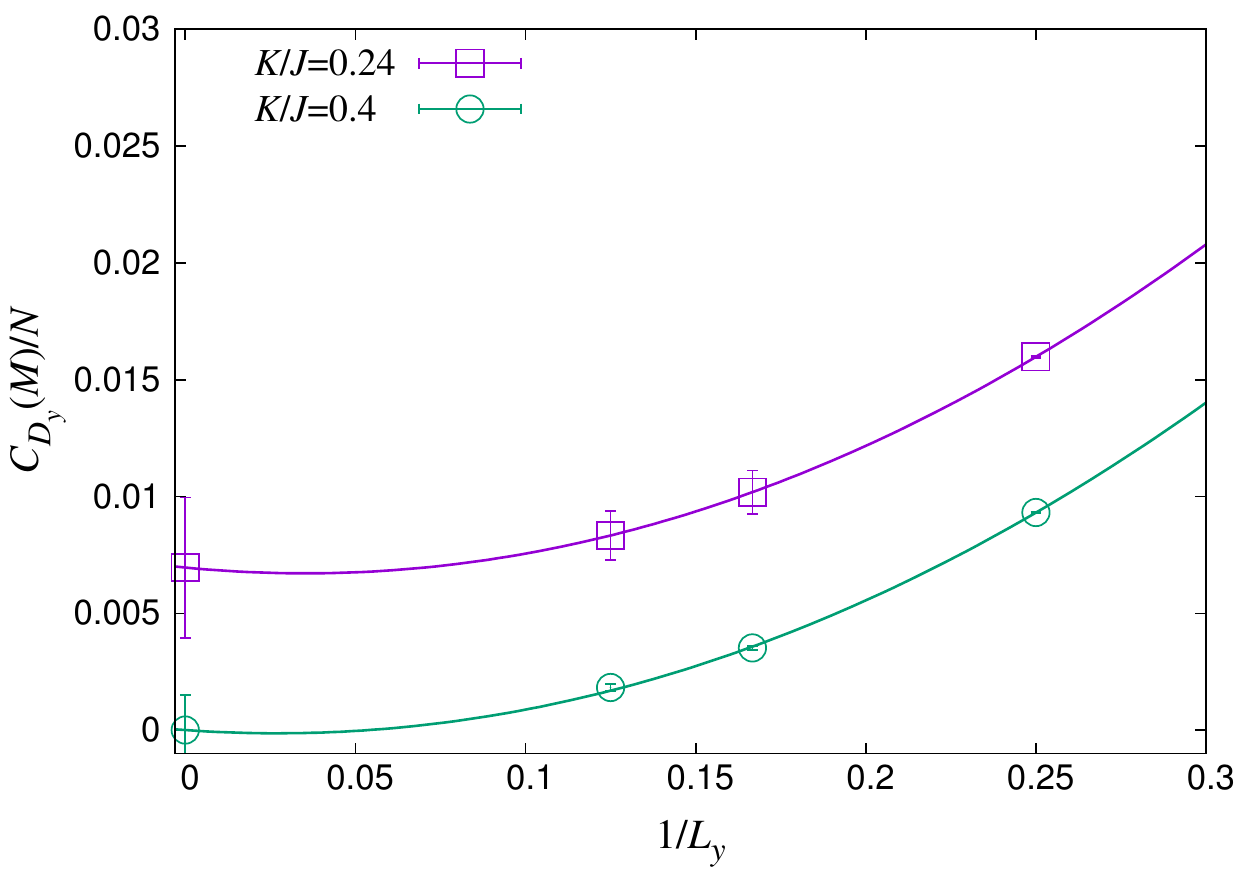}
\caption{The extrapolation of square of $y$-direction dimerized order ($C_{D_y}(M)/N$) to infinite $L_y$ with $L_x=3L_y$, where $N=L_xL_y$. Obviously, at intermediate $K/J$ region (here we plot a representative point $K/J=0.24$), there is prominent dimerized  order.}
\label{fig:DyextravsLx}
\end{figure}

In the intermediate $K/J$ region we get VBS phase, which forms orders of bond dimer, also called dimerized phase. As the original model has the threefold rotational  symmetry, there is no difference among $\hat{x}$, $\hat{y}$ or $\hat{x}+\hat{y}$ direction dimerized phase. However, in finite size DMRG calculation with cylinder geometry we have broken this symmetry and lift the degeneracy, that is why we only find $y$-direction dimerized phase survive as the dominant. Fig.~\ref{fig:Dyk_M024} shows the structure factor of $y$-direction dimer correlation for $K/J$=0.24 with $L_x=12$, $L_y=6$. The finite size scaling with fixing $L_x$ and $L_y$ ratio $L_x/L_y=3$, is showed in Fig.~\ref{fig:DyextravsLx}. It is obvious that at $K/J=0.24$, we have dimer order in thermodynamic limit.

\begin{figure}
\includegraphics[width=0.5\columnwidth]{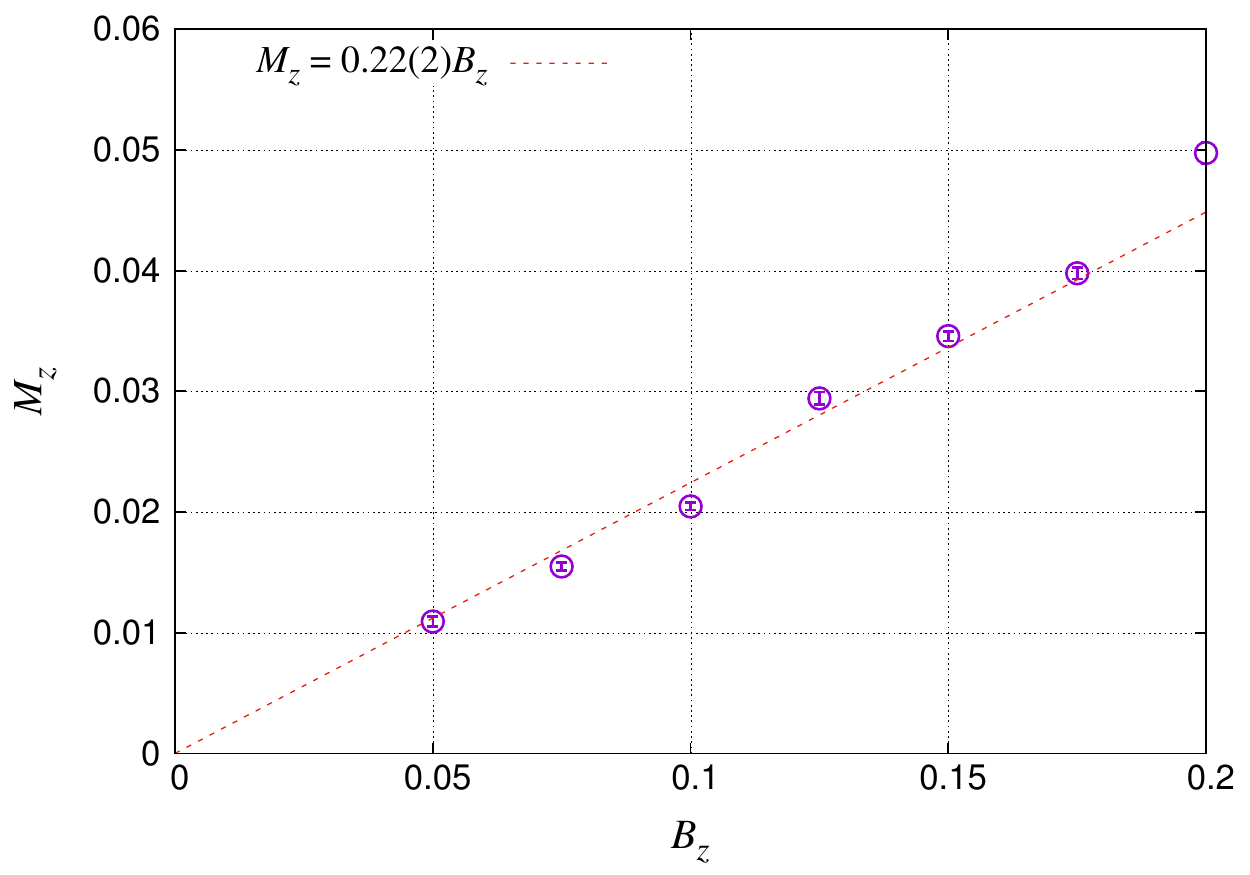}
\caption{Magnetic moment density with magnetic fields. The calculation is performed at $N_x=12$, $N_y=6$,  $\gamma=0$, $K/J$=0.8, in SFS phase. To reduce finite size effect, $y$-direction twist boundaries are used. The circle points with errorbars are DMRG calculated magnetic moment density data. The errorbars are estimated from the extrapolation of magnetic moment density to zero truncation error. The black dashed line is the linear fitting, the slope gives the static spin susceptibility to be  $\chi=0.22(2)$. }
\label{static_suscep}
\end{figure}
\subsection{SFS phase}
In $K/J>0.3$  region, we have SFS phase. We have showed real space decay of various correlations in the main text. In SFS phase, there is no spin or bond order from different structure factors  and the real space decay of all correlations indicate long range correlation.
In addition, we measured static spin susceptibility by applying $z$-direction magnetic fields with a coupling $-B_z \sum_i S_i^z$ and computed $M_z=\frac 1 N \sum_i \langle S_i^z \rangle$. As we know $\chi=\frac {\partial M_z}{\partial B_z}$, the linear fitting of the magnetic moment density with magnetic fields will give us an estimation of the static spin susceptibility. Fig~\ref{static_suscep} is such an estimation, and gives a finite value of static spin susceptibility $\chi = 0.22(2)$ (we have set $J=1$ as the energy unit). Restoring units, we have $\chi=0.22(g\mu_B)^2/J$. According to a band theory of half-filled free spinon on the triangular lattice, the static spin susceptibility is
$(g\mu_B)^2 N(0)/4$, where density of states  at Fermi surface is about $N(0)\approx\frac {1} {1.65\pi t}$ with $t$ the spinon hopping parameter.  In combination, we can also make an estimation of the hopping strength $t$ to be $0.22J$. The definitive evidence of SFS phase comes from the observation of $2k_F$ surface in the spin structure factor, which we have discussed in detail in the main text.

\begin{figure}[h!]
\includegraphics[width=0.5\columnwidth]{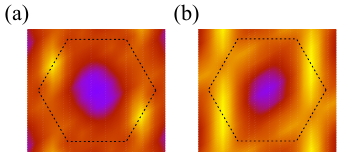}
\caption{(a) $y$-direction dimer structure factor for $L_x=12$, $L_y=6$ at $K/J=0.8$ and $\gamma=0$, in SFS phase. (b) $xy$-direction dimer structure factor with same parameters.
Dash line denotes Brillouin zone of triangular lattice.}
\label{fig:DkK08}
\end{figure}

In addition to the spin correlation function discussed in the main text, it is interesting to consider the dimer correlation function $C_{D_b}$ defined after Eq.~\eqref{eq:spincorr}. In Fig.~\ref{fig:DkK08}, we show the Fourier transform of this correlator for $b=y$ (Fig.~\ref{fig:DkK08}(a)) and $b=xy$ (Fig.~\ref{fig:DkK08}(b)). Note that the $D_y$ correlator shows a broad peak at the {\it same} M point as in Fig.~\ref{fig:Dyk_M024} and looks like a broadened version  of it. The $D_{xy}$ correlator should be equivalent by 60 degrees rotation but it appears more smeared in Fig.~\ref{fig:DkK08}(b) due to finite size effect. It is thus very interesting that the SFS state retains strong remnants of the VBS phase in the dimer correlators. Perhaps the SFS phase can be thought of as the quantum melted version of the VBS phase. We also point out that if  pair density wave (PDW) order is responsible for a gap on the Fermi surface along the $\Gamma$ to M  direction as we suggested in the main text, the wave vector of the PDW is $2k_F$ in that direction. We expect to form a charge density wave or bond density wave with wave vector $4k_F$ in the same direction as a composite order parameter which may exhibit a slow power law decay. It turns out that up to Umklapp, this wave-vector is very close to the M point. So the peak structure in the dimer correlator shown in Fig.~\ref{fig:DkK08} can be regarded as a possible signature of PDW order or a Z2 spin liquid. This is only a speculative idea at this point and much more work will clearly be needed in the future.

\begin{figure}[h!]
\includegraphics[width=0.5\columnwidth]{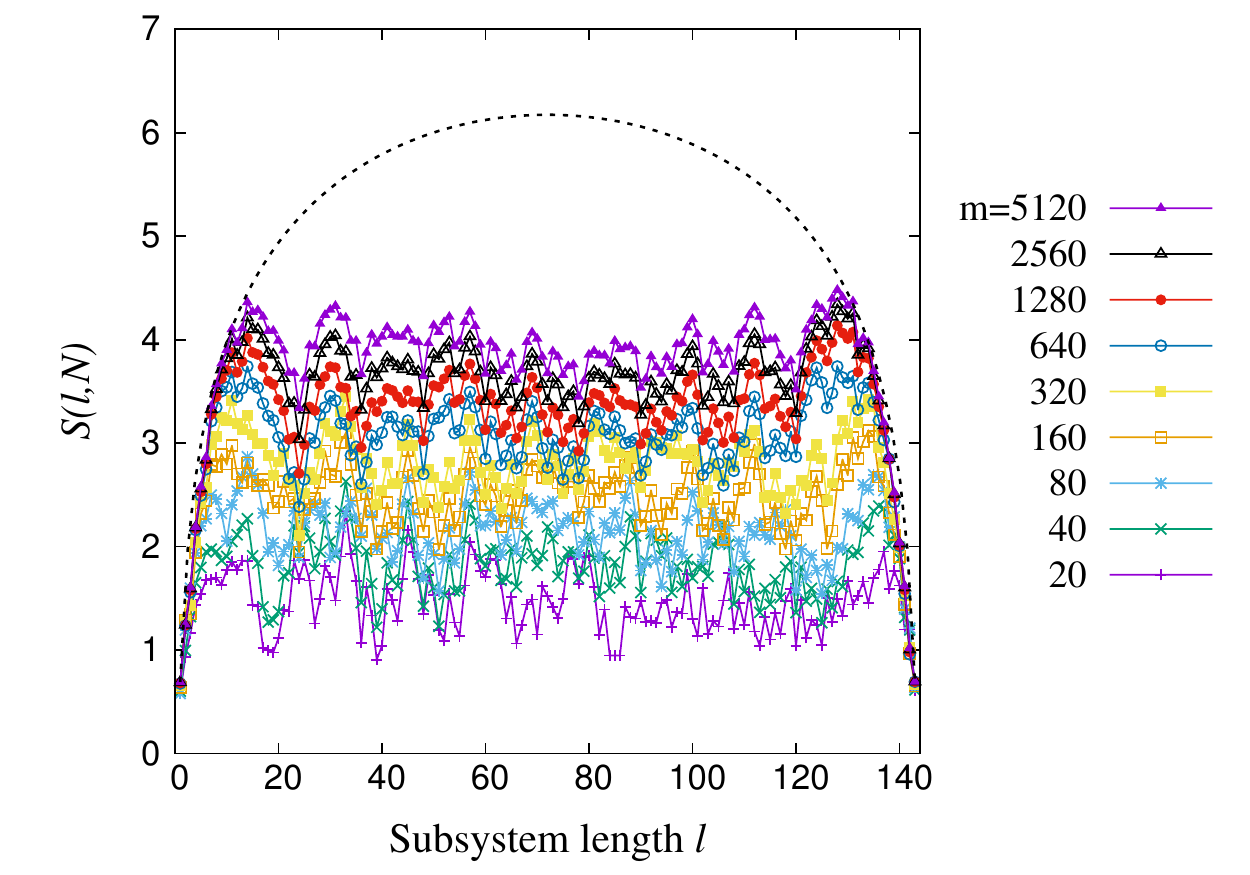}
\caption{Subsystem entanglement entropy for $L_x=24$, $L_y=6$ system at $K/J=0.8$ in SFS phase. The horizontal axis is for subsystem length and the total length is $N=L_xL_y$. Different lines correspond to entanglement entropy measured when different number of states are kept. Here we see even at $m=5120$, the entanglement entropy is still not converged especially as the subsystem extends to the center region of the system.}
\label{fig:Ly6SvN}
\end{figure}

Finally, another numerical signature of entering SFS phase is that the simulation become harder and harder to converge. In Fig.~\ref{fig:Ly6SvN}, we show the subsystem entanglement entropy for $L_x=24$, $L_y=6$ systems in SFS phase. The convergence is very slow even we increase bond dimension exponentially. At $m=5120$, the truncation error is already at the level of $10^{-5}$, but the entanglement entropy still increase slowly, especially at the center region of the system. Although the full system is not conformally invariant, we will proceed with the following analysis.
For one-dimensional gapless state with conformally invariant correlation functions in space-time, the subsystem entanglement entropy of cylinder geometry system is~\cite{Calabrese2004} 
\begin{equation}
S(l,N)=\frac c 6 \log \left( \frac N {\pi} \sin \frac {\pi l} {N} \right) + A
\label{eq:centralcharge}
\end{equation}
where $l$ is the  subsystem length, $N$ is the total length, $A$ is a constant and $c$ is the effective central charge which measures the number of gapless modes directly. From Fig~\ref{fig:Ly6SvN}, we find for small subsystem size, the entanglement is already converged, if we use this part to fit the central charge, we will get $c\approx 8.6$.
The black dashed line shows the function in Eq.\eqref{eq:centralcharge} with this central charge value. If we do similar fitting for systems with different widths, we will find the central charge increase roughly linearly with the width of the cylinder, that is to say, the number of gapless mode increases linearly with the width of cylinder, which is consistent with the SFS phase. We note that in the U(1) SFS in a six wide system, there are five bands that cross the Fermi surface giving ten gapless modes. However, one mode should be removed due to coupling to the U(1) gauge field which enforces a constraint and we expect $c=9$. On the other hand, for a Z2 SFS, the gauge photon is gapped due to spinon pairing and if a pairing gap exists along the $\Gamma$ to M direction as we suggested in the main text, we expect four bands to cross the Fermi surface and $c=8$. Unfortunately due to the slow convergence for subsystems near the middle of  the sample, we cannot determine the central charge with sufficient accuracy to  distinguish between the two cases.

Although from the entanglement entropy point of view, the convergence is slow,  the $2k_F$ peaks positions converge rather quickly,  as does the spin structure factor itself when increasing bond dimensions as showed in Fig.~4 of main text.

\subsection{Phase transition}
\begin{figure}[h!]
\includegraphics[width=0.5\columnwidth]{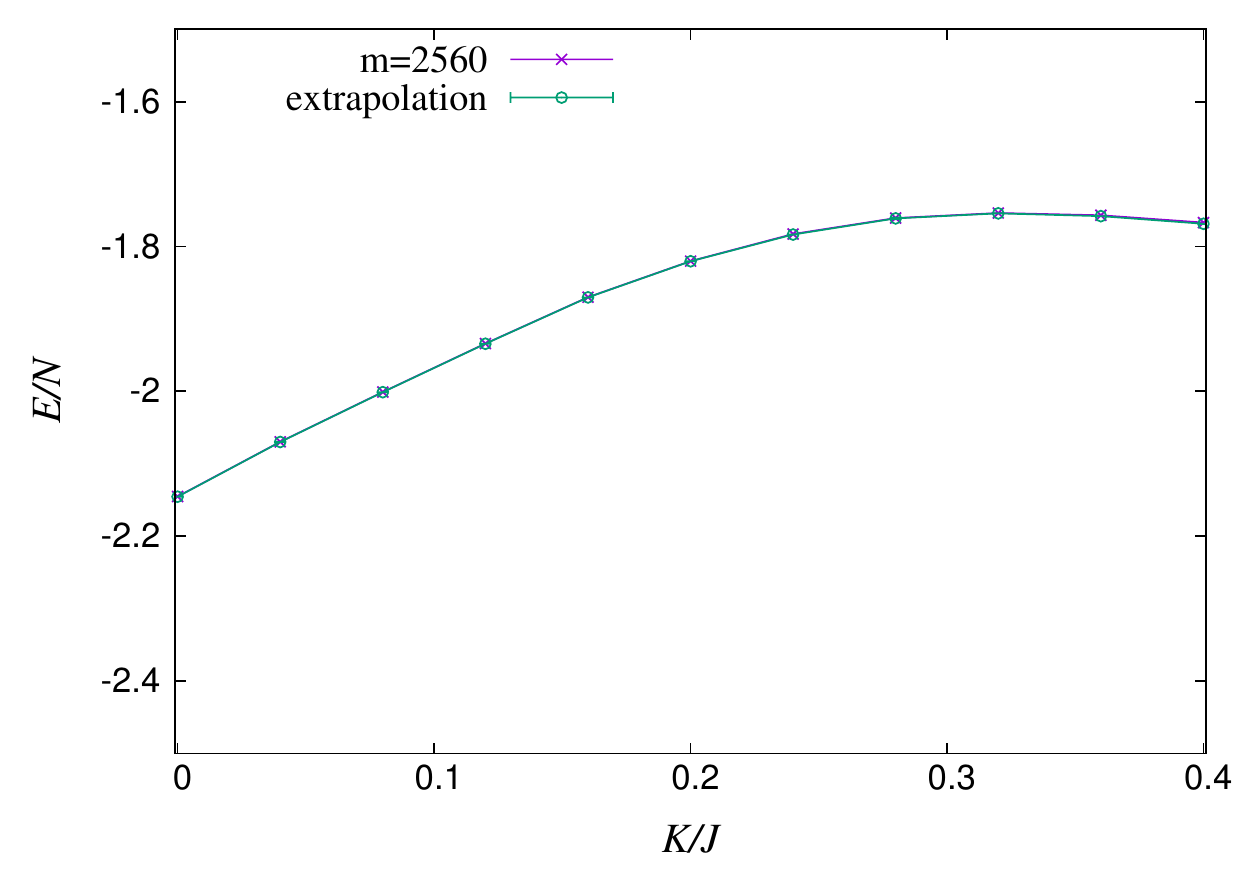}
\caption{Energy persite versus $K/J$  for $L_x=12$, $L_y=6$ with $\gamma=0$. No signature of singularity of first derivative is found. Where "extrapolation" denotes values got from extrapolation to zero truncation error.}
\label{fig:ene_vs_K}
\end{figure}

Being limited by the computation resources, we only focus on different phases here and leave the study of details of phase transitions to future works. However, from the total energy data of different $K/J$ values we have, it seems both phase transitions of AFM to VBS and VBS to SFS are continuous. Fig.~\ref{fig:ene_vs_K} shows the energy density versus $K/J$ for $L_x=12$, $L_y=6$ with $\gamma=0$ systems. There is not any singularity of first derivative, although confidence of the information is limited as we need to also consider finite size broadening.

\section{The role of anisotropy}
\begin{figure}[h!]
\includegraphics[width=0.5\columnwidth]{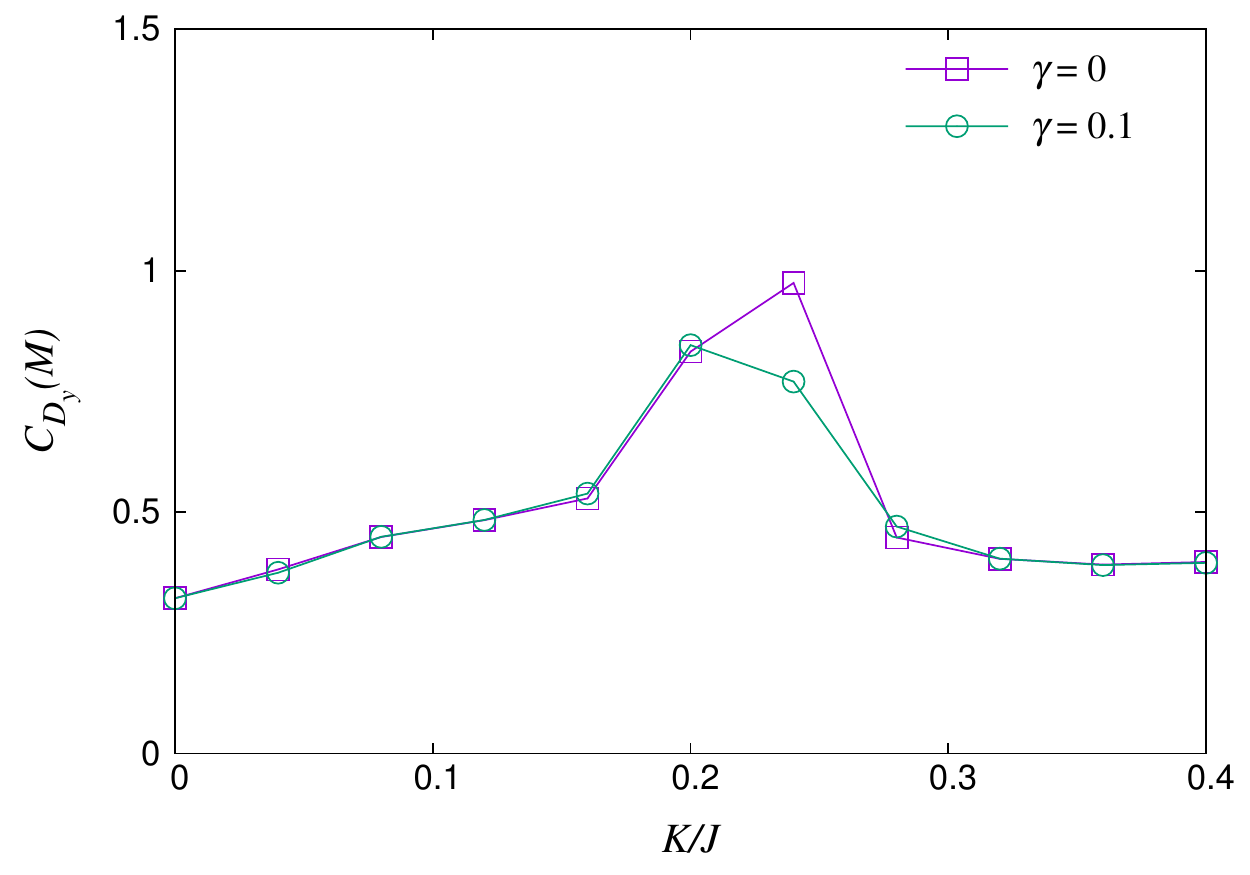}
\caption{$y$-direction dimer structure at $M$ point  versus $K/J$  for $L_x=12$, $L_y=6$ with $\gamma=0$ and $\gamma=0.1$.}
\label{fig:Dimery_vs_K}
\end{figure}

The $\gamma$ in 1T-TaS$_2$ is very close to zero and consider $\gamma=0$ is enough, but for a complete model study, we will also discuss the small anisotropy case. For a small anisotropy the overall phase diagram is the same as isotropic ($\gamma=0$) case. From the available data we have, small anisotropy may shrink the VBS phase a little bit.  In Fig.~\ref{fig:Dimery_vs_K}, we show the y-direction dimer correlation structure factor at $M$ point versus $K/J$ for $L_x=12$, $L_y=6$ with $\gamma=0$ and $\gamma=0.1$. Where a suppressing of y-direction dimer order  in small anisotropy case ($\gamma=0.1$) is found. 

\section{Comparing with earlier studied model}
Comparing our model with four-spin ring exchange model studied before~\cite{Motrunich2005,Sheng2009,Block2011},  we find if we expand the permutation operators in their model explicitly to spin operators, then the relations between our model (considering $\gamma=0$ isotropic case, with free parameters $J$ and $K$) and their model (with parameters $J_2$ and $J_4$ ) will be $J=2J_2+5J_4$, $K=4J_4$, and an extra next nearest neighbor (nnn) two spin exchange term (with exchange strength $J_4$) is  in their model while absent in ours.  Our results will be a good reference to discuss the role of nnn term and four spin ring exchange term. On the one hand,  we find SFS phase without nnn term while both VMC and DMRG simulations find SFS phase with nnn term. That means nnn term is not essential to make SFS phase, but four spin ring exchange terms should be. On the other hand nnn term may be important to make other QSL phase. For example, without four-spin ring-exchange term and considering $J_1$-$J_2$ model on the triangular lattice, several groups, with coupled-cluster~\cite{Li2015}, DMRG~\cite{Zhu2015,Hu2015}, or Gutzwiller-projected methods~\cite{Ryui2014,Iqbal2016}, find QSL phase in medium $J_2/J_1$ region (for example in Ref.~\cite{Zhu2015,Hu2015,Iqbal2016}, $0.08<J_2/J_1<0.16$), although controversies still exist in whether it is gapped or nodal QSL.

\end{document}